\documentclass{aa}  

\usepackage{graphicx}
\usepackage{txfonts}
\usepackage{float}

\usepackage[colorlinks=true,urlcolor=blue,citecolor=blue,linkcolor=blue]{hyperref}
\def\Halpha{\mbox{H\hspace{0.1ex}$\alpha$}}
\def\Hbeta{\mbox{H\hspace{0.1ex}$\beta$}}
\def\Hepsilon{\mbox{H\hspace{0.1ex}$\varepsilon$}}

\begin{document}

   \title{Formation of \Hepsilon{} in the solar atmosphere}

   \author{K. Krikova\inst{1,2}, T. M. D. Pereira \inst{1,2}, L. H. M. Rouppe van der Voort \inst{1,2}}

   \institute{Rosseland Centre for Solar Physics, University of Oslo, PO Box 1029 Blindern, 0315 Oslo, Norway \\
   \email{kilian.krikova@astro.uio.no}
    \and
    Institute of Theoretical Astrophysics, University of Oslo, PO Box 1029, Blindern 0315, Oslo, Norway}



  \abstract
   {In the solar spectrum, the Balmer series line \Hepsilon{} is a weak blend on the wing of \ion{Ca}{ii}~H. Recent high-resolution \Hepsilon{} spectroheliograms reveal a reversed granulation pattern and in some cases, even unique structures. It is apparent that \Hepsilon{} could potentially be a useful diagnostic tool for the lower solar atmosphere.}
   {Our aim is to understand how \Hepsilon{} is formed in the quiet Sun. In particular, we consider the particular physical mechanism that sets its source function and extinction, how it is formed in different solar structures, and why it is sometimes observed in emission. }
   {We used a 3D radiative  magnetohydrodynamic (MHD) simulation that accounts for non-equilibrium hydrogen ionization, run with the \emph{Bifrost} code. To synthesize \Hepsilon{} and \ion{Ca}{ii}~H spectra, we made use of the RH code, which was modified to take into account the non-equilibrium hydrogen ionization. To determine the dominant terms in the \Hepsilon{} source function, we adopted a multi-level description of the source function. Making use of the synthetic spectra and simulation, we studied the contribution function to the relative line absorption or emission and compared it with atmospheric quantities at different locations.}
   {Our multi-level source function description suggests that \Hepsilon{} source function is dominated by interlocking, with the dominant interlocking transition being through the ground level, populating the upper level of \Hepsilon{} via the Lyman series. This makes the \Hepsilon{} source function partly sensitive to temperature. The \Hepsilon{} extinction is set by \mbox{Lyman-\hspace{0.1ex}$\alpha$}. In some cases, this temperature dependence gives rise to \Hepsilon{} emission, indicating heating. The typical absorption profiles show reversed granulation and the \Hepsilon{} line core reflects mostly the \ion{Ca}{ii}~H background radiation.}
   {Synthetic \Hepsilon{} spectra can reproduce quiet Sun observations quite well. High-resolution observations reveal that \Hepsilon{} is not just a weak absorption line. Regions with \Hepsilon{} in emission are especially interesting to detect small-scale heating events in the lower solar atmosphere, such as Ellerman bombs. Thus, \Hepsilon{} can  be an important new diagnostic tool for studies of heating in the solar atmosphere, augmenting the diagnostic potential of \ion{Ca}{ii}~H when observed simultaneously.}


   \maketitle


\section{Introduction}

The \Hepsilon{} spectral line is one of the lesser known members of the Balmer hydrogen series associated with the transition from energy level $n=7$ to $n=2$, at $\lambda_0=397.1202$~nm in vacuum. In the solar spectrum, \Hepsilon{} is a weak feature blended on the red wing of the \ion{Ca}{ii}~H resonance line, one of the strongest lines in the solar spectrum. Understanding the formation of \Hepsilon{} holds an intrinsic value, but it is even more relevant in the context of simultaneous observations  with \ion{Ca}{ii}~H, a widely used line to study the solar chromosphere. Combining both lines has the potential to offer additional information about the dynamics of the solar atmosphere.


 Overall, \Hepsilon{} is rarely studied in the solar context, but it has received more attention in other stars, where it is used as an indicator of chromospheric activity \citep{Montes1995, Montes1996}. Only a handful of studies have focused on modeling \citep{Ayres1975, Bayazitov1991} and observing \Hepsilon{} \citep{Evershed1930, Turova1994} in the Sun. Observations of \Hepsilon{}  are either connected to studies of solar flares \citep{Rolli1995, Rolli1998a, Rolli1998b} or prominences \citep{Nikaidou1982, Anan2017, Zapi2022}. What all of these studies have in common is that they highlight cases where \Hepsilon{} is seen in emission relative to the \ion{Ca}{ii}~H background. The most archetypical case of \Hepsilon{} emission can be found in the K giant Arcturus \citep{Wilson1938, Wellemann1940, Popper1956, Ayres1975} or late-type stars such as K and M giants \citep{Wilson1957}. 
 
 The work of \citet{Ayres1975} is one of the first studies of the formation of \Hepsilon{} in the solar atmosphere, but it was limited to 1D plane-parallel atmospheres and disk-averaged solar observations. One of the main conclusions drawn from this study is that the \Hepsilon{} source function is dominated by the Balmer continuum radiation field, both in the Sun and Arcturus. In the case of Arcturus, the authors argue that \Hepsilon{} appears in emission due to a different column mass structure and a stronger sensitivity to the chromospheric temperature increase. As we show later in the present work, high spatial resolution solar observations taken at the \Hepsilon{} line core show a reversed granulation pattern. If \Hepsilon{} is formed under optically-thick conditions, this is difficult to reconcile with its source function being dominated by the Balmer continuum, since, in that case, we would expect to observe photospheric granulation in the \Hepsilon{} line core. This puzzling aspect has prompted us to revisit the formation of \Hepsilon{}.

 Additional motivation for this work came from a recent study by \citet{Joshi2020}, who show a much higher occurrence of quiet Sun Ellerman bombs (QSEBs) in \Hbeta{} than in \Halpha{} observations. \citet{Joshi2020} raises the question of whether \Hepsilon{} could be an even better diagnostic tool for detecting QSEBs. Because of its shorter wavelength, \Hepsilon{} has the potential to probe photospheric reconnection at even higher spatial resolution and enhanced intensity contrast, and it is less affected by extinction from the overlying chromospheric canopy fibrils.

In this study, we want to understand the formation of \Hepsilon{} in the dynamic solar atmosphere, using a state-of-the-art 3D radiative magneto-hydrodynamic (MHD) simulations in combination with non-LTE radiative transfer. The outline of the paper is as follows. In Sect. \ref{sec:Observations} we describe recent observations of the  \ion{Ca}{ii}~H  and \Hepsilon{} lines. Section \ref{sec:Methods} details our forward-modeling methods aimed at understanding the formation of \Hepsilon{} in a dynamic atmosphere. Our results are presented in Section \ref{sec:Results}, where we detail the formation of \Hepsilon{} for different structures seen in observations and investigate the influence of Balmer continuum line blanketing, hydrogen non-equilibrium ionization, and 3D radiative transfer on the formation of \Hepsilon{}. We  present our discussion in Sect. \ref{sec:Discussion}, followed by our concluding remarks in Sect.~\ref{sec:Conclusion}.

\section{Observations}
\label{sec:Observations}

To compare our synthetic spectra with solar observations, we made use of data taken with the CHROMIS instrument installed at the Swedish 1-m Solar Telescope \citep[SST,][]{Scharmer2003} located at La Palma.
CHROMIS is a Fabry-P{\'e}rot filtergraph based on the design by \citet{Scharmer2006} that  is capable of fast wavelength sampling. We observed the \ion{Ca}{ii}~H line together with \Hepsilon{} at 47 wavelength positions with a cadence of 13~s. 

CHROMIS has a narrowband spectral transmission profile width of $0.012$~nm. We covered a spectral range from $396.63$~nm to $397.09$~nm. Spectral sampling was performed with a critical sampling of $0.006$~nm steps within $\pm 0.072$~nm around the \ion{Ca}{ii}~H line core and between an offset of $+0.132$ to $+0.168$~nm relative to the \ion{Ca}{ii}~H line core, covering the \Hepsilon{} line. The \Hepsilon{} line is located at $+0.156$~nm redward of the \ion{Ca}{ii}~H line core. The sampling positions outside of the ranges mentioned above were taken with $0.012$~nm steps. An additional wavelength position at $400$~nm was used to sample the continuum. 
CHROMIS includes a wide band (WB) channel that is used for image restoration. The WB imaging filter has a center wavelength of 395~nm, a full width at half maximum (FWHM)\ of 1.32~nm, and displays a photospheric scene. 

CHROMIS data were processed using the SSTRED data processing pipeline,
\citep{2021A&A...653A..68L}, 
which includes multi-object multi-frame blind deconvolution 
\citep[MOMFBD,][]{2005SoPh..228..191V} 
image restoration. 
Here, we analyzed spectral scans that were recorded under excellent seeing conditions and the spatial resolution is close to the diffraction limit of the telescope ($\lambda/D = 0\farcs08$ at the wavelength of \ion{Ca}{ii}~H for the $D=0.97$~m clear aperture of the SST). 
The SST adaptive optics wavefront sensor measured the seeing quality
\citep[see][]{2019A&A...626A..55S}. 
The Fried's parameter $r_0$ for the ground-layer seeing peaked above 50~cm during the best seeing moments. 
CHROMIS has a pixel size of $0\farcs0375$.

The targets of our CHROMIS observations were a quiet Sun and a pore region. The quiet Sun observation was taken on June 22, 2021 at 15:28~UT near the disk center at (x, y) = ($12\arcsec$, $14\arcsec$). The pore observation was taken on June 28, 2022 starting at 08:30~UT close to AR 13040 at (x, y) = ($232\arcsec$, $-220\arcsec$) and is part of a longer time series. Two snapshots of this series are shown in this work. The first snapshot was taken around 08:49:45~UT and the second around 09:17:52~UT. For the purposes of our work, we only show a small part of the total field of view with a spatial extent of $\approx$24~Mm and $\approx$17~Mm (for the 09:17:52~UT snapshot).

\section{Methods}
\label{sec:Methods}
\subsection{Model atmospheres}
\label{Model Atmospheres}
We used two different model atmospheres in our study. First, the semi-empirical model C from \citet{Fontenla1993} which we will refer to as the FALC model. The 1D time-independent FALC atmosphere was constructed to match temporal and spatial averaged spectra in the UV. This ``simple'' atmosphere gives a good starting point to understand the formation of \Hepsilon{} and compare synthetic profiles to temporal and spatial averaged observations of the Sun, such as observations from the Fourier transform spectrometer (FTS) atlas \citep{Kurucz1984}.

The second model atmosphere we use is the publicly available Bifrost model from \citet{Carlsson2016}. Bifrost is a 3D radiation magneto-hydrodynamic simulation code that solves the resistive MHD equation on a Cartesian grid including various important physical effects in the solar atmosphere \citep[for more details, see][]{Gudiksen2011}. One important physical ingredient is hydrogen non-equilibrium ionization (HNEI), influencing the hydrogen level populations and, therefore, the formation of \Hepsilon{}. The Cartesian grid of the numerical model consists of $504 \times 504 \times 496$ grid points, spanning a computational box of $24 \times 24 \times 14.4$~Mm. The simulation consists of multiple snapshots in time. We used snapshot 385 at simulation time $t=3850$~s, which is the first published snapshot with non-equilibrium hydrogen ionization.

\subsection{Model atoms}
\label{Model Atoms}

The \Hepsilon{} line is a weak line blended in the strong \ion{Ca}{ii}~H wing.
Therefore, we have to properly model the \ion{Ca}{ii}~H line with partial frequency
redistribution (PRD) over the line profile as the inner wings of \ion{Ca}{ii}~H are
affected by PRD at the \Hepsilon{} wavelength. To model \ion{Ca}{ii}~H, we used the
five-level plus continuum Ca model atom from \citet{Carlsson2012}. We modeled the
infrared triplet of Ca II (at $849.8$, $854.2$, and $866.2$~nm) with the assumption of complete frequency redistribution (CRD), as well as the Ca II K line. The resonance doublet and infrared triplet of Ca II are affected by the cross-redistribution (XRD) of photons between these two spectral lines, generally
known as Raman scattering. Therefore, XRD will affect the inner wings of the \ion{Ca}{ii}~H
and K line mapping heights where the Ca triplet lines are formed. These give a different escape route of \ion{Ca}{ii}~H and K photons and will decrease the inner wing
intensity of the lines. The influence of XRD effects in \ion{Ca}{ii} was studied in some detail by \citet{Bjorgen2018}. We tested the influence of XRD on the
\ion{Ca}{ii}~H wing intensity at the wavelength where \Hepsilon{} is located and found minor intensity changes if we treat the doublet and triplet lines in XRD. Consequently,
we only model \ion{Ca}{ii}~H in PRD and treat all the other lines in CRD to decrease
the computational time. We used the approximation of angle-dependent PRD of \citet{Leenaarts2012b}.

To construct an efficient hydrogen model atom to synthesize \Hepsilon,{} we started with a nineteen-level plus continuum model atom and collapsed collectively levels starting from the uppermost hydrogen energy level down to the upper energy level of \Hepsilon{} ($n=7$). The goal was to find the smallest hydrogen model atom, which is computationally the most efficient and models the \Hepsilon~line best, as compared with the nineteen-level hydrogen atom. We found that an eight-level plus continuum hydrogen atom, where the eighth-level consists of the collapsed levels from $n=19$ to $n=8$, provides a good approximation of the \Hepsilon{} line when compared with the nineteen-level atom. Other collapsed model atoms seem to model the \Hepsilon{} line less accurately (although, generally, the difference is small), by representing the recombination cascades of hydrogen less correctly. We tested if the assumption of PRD for other hydrogen transitions has an influence on the emergent \Hepsilon{} line profile. Tests show that the influence of PRD in \Hepsilon\ is negligible and we therefore  treated all hydrogen transitions in CRD. To summarize, we chose to model \Hepsilon{} with a collapsed eight-level plus continuum atom where \Hepsilon{} is calculated in CRD as well as all other hydrogen transitions. 

\subsection{Total source function}
\label{Total Source function}

The total source function, $S^\mathrm{t}_\nu$, at the wavelength of \Hepsilon{} can be written as an expression of the line source function, $S^\mathrm{l}_{\nu}$, and background source function, $S^\mathrm{b}_\nu$,
\begin{equation}\label{eq:total_source_function}
     S^\mathrm{t}_\nu = \delta_\nu \, S^\mathrm{l}_{\nu} + \left(1 - \delta_\nu \right) \, S^\mathrm{b}_\nu,
\end{equation} 
where
\begin{equation}\label{eq:relative_opacity}
     \delta_\nu = \frac{\chi^\mathrm{l}_\nu}{\left(\chi^\mathrm{l}_\nu + \chi^\mathrm{b}_\nu \right)},
\end{equation} 
which is the relative line extinction. The relative line extinction gives the probability per total extinction that a photon is destroyed by a line process.
The background source function, $S^\mathrm{b}_\nu$, contains all  background emissivity and extinction contributions (\ion{Ca}{ii}~H, continuum and others) at the \Hepsilon{} wavelength not related to the \Hepsilon{} transition.

\subsection{Radiative transfer}
\label{Radiative Transfer}

We synthesized the Ca II  H line with overlapping \Hepsilon{} wavelength points using the RH code \citep{Uitenbroek2001} and the 1.5D version \citep{Pereira2015}. This code solves the non-LTE radiative transfer problem with overlapping wavelength points for bound-bound and bound-free transitions. Forward modeling with the RH 1.5D code will treat every vertical column of a 3D model atmosphere as an independent plane-parallel 1D atmosphere (1.5D approximation) with no horizontal transfer of radiation. Because of this shortcoming, we went on to use the 3D branch of RH to synthesize \ion{Ca}{ii}~H and \Hepsilon{} to check for 3D effects in a few regions.

To study the formation of the \Hepsilon{} spectral line, we need to distinguish between the region of formation of the emergent intensity; in our case, this is the \ion{Ca}{ii}~H wing intensity and the region where the \Hepsilon{} line depression is formed. As we are dealing with a ``weak'' line blend in a strong background continuum, the \ion{Ca}{ii}~H line, the use of contribution function (CF) to intensity can be misleading \citep{Magain1986} and can give a misleading impression of where the line depression is formed. In the case of faint or weak spectral lines, one has to distinguish between the origin of the continuum and the origin of the region where the line depression is formed. Otherwise, this could lead to a wrong interpretation where the emergent intensity of a spectral line is formed \citep{deJager1952, Gurtovenko1974}. \citet{Magain1986} gives a derivation of the relative contribution function to intensity (line depression or line emission):
\begin{equation}
    R_\nu=\frac{\left(I^\mathrm{b}_\nu - I^\mathrm{l}_\nu \right)}{I^\mathrm{b}_\nu},
\end{equation}
and expresses the transfer equation for $R_\nu$, identifying the integrand of the formal solution as the CF to relative line depression \citep[for a detailed derivation, see,][]{Magain1986}. Also, $I^\mathrm{b}_\nu$ is the background (continuum) intensity and $I^\mathrm{l}_\nu$ is the intensity of the spectral line. Thus, the formal solution and relative contribution function for emergent line depression can be expressed as:
\begin{gather}
    R_\nu = \int_{z_0}^{z_1} S^\mathrm{R}_\nu \, e^{\tau^\mathrm{R}_\nu} \, \chi^\mathrm{R}_\nu \, \textrm{d}z, \label{eq:formal solution relatove} \\ 
    C_\mathrm{R}(\nu, z) = \frac{\textrm{d}I_\mathrm{R}}{dz} = S^\mathrm{R}_\nu \, e^{-\tau^\mathrm{R}_\nu} \, \chi^\mathrm{R}_\nu \label{eq:relative cf},
\end{gather}
 with $S^\mathrm{R}_\nu$ and $\chi^\mathrm{R}_\nu$ given as:
\begin{gather}\label{eq:relative source function}
    S^\mathrm{R}_\nu = \left(1 - \frac{S^\mathrm{l}_\nu}{I^\mathrm{b}_\nu} \right)/ \left(1 + \frac{\chi^\mathrm{b}_\nu}{\chi^\mathrm{l}_\nu} \frac{S^\mathrm{b}_\nu}{I^\mathrm{b}_\nu} \right), \\ 
    \chi^\mathrm{R}_\nu = \chi^\mathrm{l}_\nu + \chi^\mathrm{b}_\nu \, \frac{S^\mathrm{b}_\nu}{I^\mathrm{b}_\nu}.
\end{gather}

The relative contribution function can be simplified to
\begin{equation}\label{eq:relative_contr}
     C_\mathrm{R}(\nu, z) = \frac{\textrm{d}I_\mathrm{R}}{dz} = \chi^\mathrm{l}_\nu \, \left(1- \frac{S^\mathrm{l}_\nu}{I^\mathrm{b}_\nu} \right) \,  e^{-\tau^\mathrm{R}_\nu},
\end{equation}
which gives a more intuitive way to understand relative spectral line formation. We see a contribution to line depression or emission if $\chi^\mathrm{l}_\nu \neq 0$ or $S^\mathrm{l}_\nu \neq I^\mathrm{b}_\nu$. The former says that some absorber has to be present for a spectral line to form and the latter tells us that the photons created by the line processes should not be equal to the photons created by the background processes at the region where the line is formed. Further, the expression in the brackets of Eq. (\ref{eq:relative_contr})  defines whether there is a contribution to relative line emission $S^\mathrm{l}_\nu > I^\mathrm{b}_\nu$ or depression $S^\mathrm{l}_\nu < I^\mathrm{b}_\nu$. In our case, $I^\mathrm{b}_\nu$ represents the height-dependent \ion{Ca}{ii}~H background wing intensity.
 
\subsection{Statistical equilibrium and approximation to HNEI}
\label{sec:stat_equil}
 
In a dynamical atmosphere, such as the solar chromosphere, the statistically time-dependent state of the plasma is given by:

\begin{equation}\label{eq:kinetci_equil_eq}
     \sum_{j \neq i} n_j \, P_{ji} - n_i \sum_{j \neq i} P_{ij} = \frac{\partial n_i}{\partial t},
\end{equation}
where $n_i$, $n_j$ are the number density of particles in state $i$ or $j$. The left-hand side gives the collision term of the kinetic equilibrium equation and is known as the statistical equilibrium equation if time-dependent changes are negligible. Then, $P_{ij}$ is the total rate of transition (collisions and radiative) from state $i$ to $j$. Radiative transfer codes such as RH solve only the statistical equilibrium equation, neglecting time-dependent changes which are important for species with large second-level excitation energy, such as hydrogen and helium \citep{Carlsson2002, Golding2014}. These time-dependent changes are available from state-of-the-art radiative MHD simulations. Currently, it is not computationally feasible to solve the exact radiative transfer problem at the same time with the MHD equations. Therefore, we can use an approximation for HNEI given by
\begin{equation}\label{eq:HNEI_eq}
     \sum_{j \neq i, c} n_j \, P_{ji} - n_i \sum_{j \neq i} P_{ij} = n_{c} \, P_{ci},
\end{equation} 
solving only for the bound levels and fixing the ionized hydrogen number density $n_c$ given by the Bifrost simulation which treats hydrogen in non-equilibrium ionization. Here, $P_{ci}$ describes the transition rate out of the continuum. We implemented this approximation to HNEI into the RH code when solving the statistical equilibrium equation. 

The inherent degeneracy of solutions to the statistical equilibrium equation allows us to formulate solutions in a variety of ways \citep{White1961}. However, not all of them are convenient for the interpretation of line source functions. A particularly interesting solution is given by \citet{Jefferies1960, Jefferies1968}, written in terms of indirect transition probabilities between levels. This formulation helps to find the strongest indirect transition contributing to the interlocking term in the multi-level source function description (see Sect. \ref{sec:multi-level}). A general formal solution in terms of level ratios was already given by \citet{Rosseland1926} and in terms of indirect transition probabilities can be written as
\begin{equation}\label{eq:stat_eq_prob}
     \frac{n_u}{n_l} = \frac{\sum_k P_{lk} \, q_{ku,l}}{\sum_k P_{uk} \, q_{kl,u}},
\end{equation} 
with $P_{lk}$ the transition rate out of the lower level $l$ into individual levels $k$ and $q_{ku,l}$ the indirect transition probability that a transition starting from the level $k$ arrives at the upper level $u$ before transitioning back to the lower level $l$. The indirect transition probabilities can be calculated by solving  a set of linear equations:
\begin{equation}\label{eq:q_indirect}
     q_{ij,k} = \sum_{l \neq i} p_{il} \, q_{lj,k},
\end{equation} 
with the the transition probability $p_{il}$ given as:
\begin{equation}\label{eq:indirect_seeq}
     p_{il} = \frac{P_{il}}{\sum_{k} P_{ik}}.
\end{equation} 

\subsection{Multi-level  source function}
\label{sec:multi-level}
 
The expression for the solution of the statistical equilibrium equation given by Eq. (\ref{eq:stat_eq_prob}) can be used to express the multi-level line source, $S^\mathrm{l}_\nu$, for the complete redistribution in a more intuitive way:

\begin{equation}\label{eq:s_l}
     S^\mathrm{l}_{\nu_0} = \sigma \, \overline{J}_{\nu_0}  + \epsilon \, B_{\nu_0} (T_\mathrm{e}) + \eta \, B_{\nu_0} (T^{\star}),
\end{equation} 

where the term $\sigma \overline{J}_{\nu_0}$ describes the contribution to line photons by scattering from the profile-averaged mean radiation field $\overline{J}_{\nu_0}$,  while $\epsilon B_\nu(T_\mathrm{e})$ is the fraction of thermally created line photon following the Planck function and the last term, $\eta B_\nu(T^\star), $ is the contribution to line photons from interlocking transitions. Interlocking transitions populate the upper or lower level of a given transition indirectly via an intermediate level. Here, we express the multi-level line source function $S^\mathrm{l}_{\nu_0}$ in terms of total extinction, while in the literature \citep[e.g.,][]{Jefferies1968} the multi-level line source function is often expressed relative to scattering. The coefficients $\sigma, \epsilon$, and $\eta$ relative to the total extinction are given by:
\begin{gather}\label{eq:equation1}
     \kappa = A_{ul} + C_{ul} \left(1 - \mathrm{exp}\left(-\frac{h\nu_0}{k_\mathrm{B} T_\mathrm{e}}\right)\right) + \left(g_u \sum_{u} - g_l \sum_{l} \right),   \\ 
     \sigma = \frac{A_{ul}}{\kappa},
     \\
     \epsilon = \frac{C_{ul} \left(1 - \mathrm{exp}\left(-\frac{h\nu_0}{k_\mathrm{B} T_\mathrm{e}}\right)\right)}{\kappa},
     \\
     \eta = \frac{\left(g_u \sum_{u} - g_l \sum_{l} \right)}{\kappa},
\end{gather}
with $C_{ul}$ and $A_{ul}$ having the usual meaning of the collisional and radiative deexcitation rates. Also, $T_\mathrm{e}$ is the electron temperature. The terms $ \sum_u, \sum_l$, and $B_\nu (T^\star)$, which determine interlocking, are given as:
\begin{gather}\label{eq:equation2}
     \sum_u = \sum_{j \neq l} P_{uj} \, q_{jl,u},
     \\
     \sum_l = \sum_{j \neq u} P_{lj} \, q_{ju,l},
     \\
     B_\nu(T^\star) = \frac{2h\nu^3}{c^2} \left(\frac{g_u \sum_u}{g_l \sum_l} - 1 \right)^{-1}  \label{eq:eq_bstar}.
\end{gather}
The quantities $\sum_u, \sum_l$ are the sum of all indirect transitions that start from the upper level $u$ (lower level $l$) and populate indirectly the lower level $l$ (upper level $u$) via some intermediate transitions. These two quantities help determine which intermediate levels play a significant role in the contribution to interlocking, whereas $B_\nu(T^\star)$ represents a Planck function with a characteristic temperature $T^\star$ determined by the dominant intermediate interlocking processes. 

To determine how strong each intermediate level contributes to interlocking $B_\nu(T^\star)$ in Eq. (\ref{eq:eq_bstar}) we can further break down the expression into a sum of Planck functions each with a characteristic temperature, $T^\star$, as:
\begin{equation}\label{eq:bstar_seperate}
     B_\nu(T^\star) = \sum_i B^i_\nu(T^\star_i) \gamma_i = \sum_i B^i_\nu(T^\star_i) \frac{\beta_i}{\beta_{\mathrm{tot}}},
\end{equation} 
where $i$ is the sum over the intermediate levels and $B^i_\nu(T^\star_i)$, $\beta_i$, and $\beta_{\mathrm{tot}}$ are given by:

\begin{gather}\label{eq:bstar_terms}
     B^i_\nu(T^\star_i) = \frac{2h\nu^3}{c^2} \left(\frac{g_u \, P_{ui} \, q_{il,u}}{g_l \, P_{li} \, q_{iu,l}} - 1 \right)^{-1},
     \\
     \beta_i = g_u \, P_{ui} \, q_{il,u} - g_l \, P_{li} \, q_{iu,l},
     \\
     \beta_{\mathrm{tot}} = \sum_i g_u \, P_{ui} \, q_{il,u} - g_l \, P_{li}\, q_{iu,l},
\end{gather}
where $B^i_\nu(T^\star_i)$ describe Planck functions following characteristic temperatures $T^\star$ determined by the radiative and collisional processes coupling the upper and lower level through a loop with the intermediate level $i$. Here,  $\beta_i$ and $\beta_{\mathrm{tot}}$ are necessary to split the ratio of $g_u\sum_u/g_l\sum_l$ into several different Planck functions. Physically, $\beta_i / \beta_{\mathrm{tot}}$ describes the fractional contribution to interlocking extinction for each intermediate loop via level $i$. $\beta_i$ the contribution to interlocking extinction corrected for indirect transitions into the upper level for each individual indirect transition loop.

Finding the physical process and transitions that set the line source function in NLTE radiative transfer can be quite challenging if indirect transitions play a significant role in populating atomic levels. It is difficult to quantify how strong intermediate transitions affect the line source function, especially for model atoms with a large number of transitions. The given description of the multi-level source function, including indirect transition probabilities for the interlocking term, gives a straightforward way to quantify which transition(s) dominate the line source function.

\begin{figure*}
    \centering
    \includegraphics[width=1\textwidth]{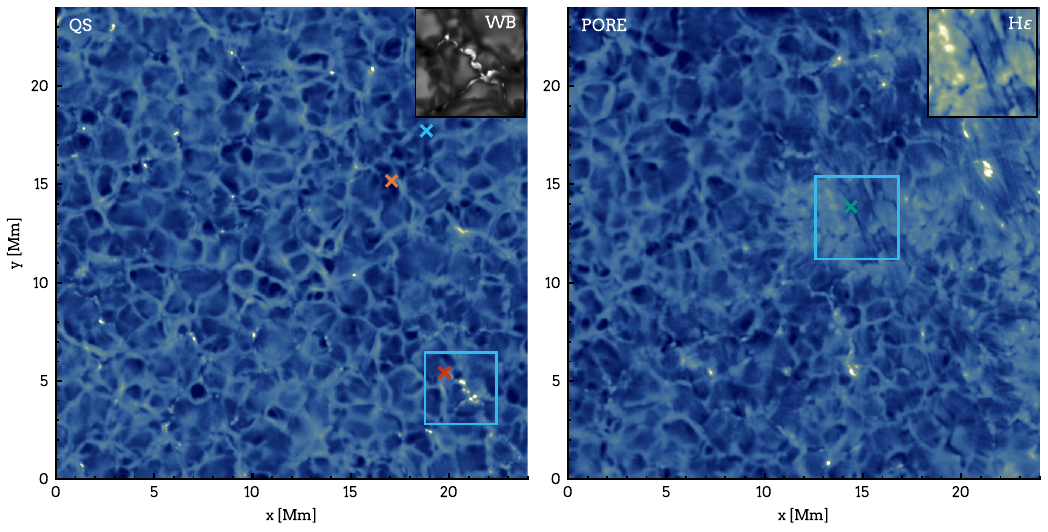}
    \caption{\Hepsilon{} line core observations of a quiet Sun and pore region. The images are histogram equalized to the synthetic \Hepsilon{} line core image in Fig.~\ref{fig:EN_SE_HNEI}. \emph{Left panel:} Quiet Sun observation with a wide band insert at the top right corner. The WB insert covers the area outlined by the light blue square and includes a few magnetic bright points. \emph{Right panel:} More active scene close to a pore region (taken at 08:49:45~UT) with a zoomed-in insert at the top right corner (the insert is more saturated to highlight dark fibrils). Crosses indicate positions of spectral profiles shown in Fig.~\ref{fig:observational_profiles} with the same color coding.}

    \label{fig:observations_Hepsilon}
\end{figure*}

\section{Results}
\label{sec:Results}

\subsection[Hepsilon observations]{\Hepsilon{} observations}
\label{Hepsilon observations}

In this section, we discuss how the morphology of \Hepsilon{} line core images can guide us to understand the formation of the spectral line. To explain the apparent structures seen in \Hepsilon{} line core images, we need to determine where we are looking in the solar atmosphere, in terms of extinction and what sets the total source function.

\begin{figure}
    \centering
    \includegraphics[width=0.5\textwidth]{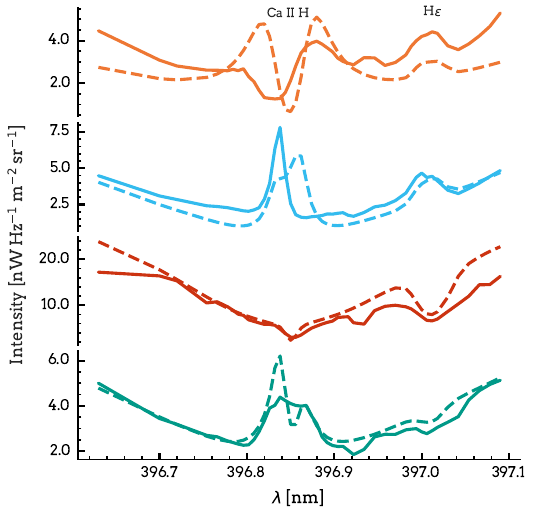}
    \caption{\ion{Ca}{ii}~H plus \Hepsilon{} spectral profiles for different solar structures indicated with crosses in Fig.~\ref{fig:observations_Hepsilon}. \emph{Solid lines:} Observations taken with the CHROMIS instrument. \emph{Dashed lines:} Degraded synthetic spectra. The \Hepsilon{} emission profiles are shown in orange and light blue. Orange synthetic profile (1.5D HNEI) is taken from Fig.~\ref{fig:EN_SE_HNEI} at $x \approx 13.6$~Mm and $y \approx 17.7$~Mm (red contours). Light blue synthetic profile (1.5D HNEI) shows the same as the one in the bottom right panel of Fig.~\ref{fig:HNEI_1D_3D_high_coldens}. The red spectral profile is taken from a magnetic bright point. The synthetic profile (3D SE) is shown in the bottom left panel of Fig.~\ref{fig:absorption_granule_magnetic}. The turquoise spectral profile is taken from a dark fibrilar structure. The synthetic profile (1.5D HNEI) is shown in the bottom right panel of Fig.~\ref{fig:absorption_granule_magnetic}.}
    \label{fig:observational_profiles}
\end{figure}

Figure \ref{fig:observations_Hepsilon} displays such \Hepsilon{} line core images. We choose these two particular regions for comparison with our synthetic spectra. Figure~\ref{fig:observational_profiles} compares synthetic degraded spectra with selected observed spectral profiles.

\begin{figure*}
    \centering
    \includegraphics[width=1\textwidth]{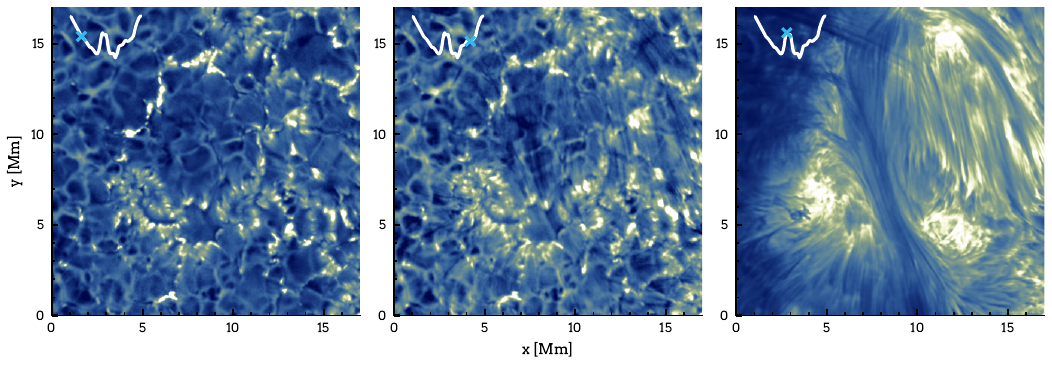}
    \caption{\Hepsilon{} and \ion{Ca}{ii}~H pore observation covering a region with stronger magnetic fields (taken at 09:17:52 UT). The wavelength positions are indicated in the top left corners on an average spectral profile with light blue crosses. \emph{Left panel:} \ion{Ca}{ii}~H blue ($-0.15$~nm) wing image, showing no chromospheric structures. \emph{Middle panel:} \Hepsilon{} line core image, showing both reversed granulation and chromospheric structures. \emph{Right panel:} \ion{Ca}{ii}~H line core image, dominated by chromospheric structures.}
    \label{fig:observations_active_sun}
\end{figure*}

We start with a discussion of where we are looking with regard to extinction in the quiet Sun case. The \Hepsilon{} line core images reveal a reversed granulation pattern. Reversed granulation originates from the mid-photosphere due to a reversal in the temperature structure, which leads to a reversal in the intensity images compared to the photospheric granulation pattern \citep{Nordlund1984, Cheung2007}. The fact that the \Hepsilon{} line core image shows reversed granulation, as well as the \ion{Ca}{ii}~H blue wing (see Fig. \ref{fig:observations_active_sun}) suggests that the layers above reversed granulation are optically thin to \Hepsilon{} radiation. As we are looking at atmospheric layers close to the temperature minimum, the \Hepsilon{} extinction should be significantly reduced due to the temperature sensitivity of the hydrogen $n=2$ population. A forward synthesis of \Hepsilon{} from the Bifrost simulation would help clarify which observed structures (if any) could be optically thick to \Hepsilon{} radiation. Furthermore, as no chromospheric canopy structure (the fibrilar chromosphere that is characteristic for a chromospheric line such as \Halpha{}) is apparent in \Hepsilon{} quiet Sun images, the chromosphere is mostly transparent, suggesting a mid-photospheric origin of \Hepsilon{}. 

The pore region shows a slightly different picture. Here, we see some imprints of dark fibrilar structures. This hints at a chromospheric origin for such fibrilar structures observed in \Hepsilon{} (see top-right insert in the right panel of Fig.~\ref{fig:observations_Hepsilon}). To illustrate that \Hepsilon{} can show some chromospheric structures, we show the \ion{Ca}{ii}~H blue wing, \Hepsilon{}, and \ion{Ca}{ii}~H line core images side-by-side in Fig.~\ref{fig:observations_active_sun}. The blue wing \ion{Ca}{ii}~H blue image shows only reversed granulation, whereas the \Hepsilon{} image shows fine imprints of dark fibrils. These dark fibrils are connected to chromospheric fibrilar structures observed in the \ion{Ca}{ii}~H line core image, illustrating a chromospheric contribution in the \Hepsilon{} images. Not all dark fibrilar structures seem to be optically thick. Through some fibrilar structures, we observe the background with reversed granulation. In more active solar regions, the \Hepsilon{} extinction appears to increase at higher layers, making the chromosphere opaque. We note the images shown were taken during excellent seeing conditions, making the thin dark fibrilar structures clearly visible. 

Next, we address the \Hepsilon{} source function. \citet{Ayres1975} suggested that the \Hepsilon{} source function is dominated by the Balmer continuum radiation field. This would result in apparent images resembling granulation if \Hepsilon{} is formed under optically-thick conditions, as the Balmer continuum is highly scattering and the thermalization depth lies in the photosphere. Nevertheless, we see reversed granulation in \Hepsilon{} images, which suggests most radiation is coming from the \ion{Ca}{ii}~H background intensity and not \Hepsilon{} itself. Detailed radiative transfer calculations can reveal which hydrogen atomic transition dominates the \Hepsilon{} source function.

The conclusion we can draw from the \Hepsilon{} observations helps to narrow down the question, how the \Hepsilon{} line is formed in the solar atmosphere and gives some restriction on the total source function (Eq. \ref{eq:total_source_function}). The total source function $S_\nu^\mathrm{t}$ has to be sensitive to temperature but it is a linear combination of the line source function, $S^\mathrm{l}_{\nu_0}$, and the background source function, $S^\mathrm{b}_\nu$, weighted by their relative extinctions. This leaves us with two options. Either the \ion{Ca}{ii}~H extinction dominates and  $S_\nu^\mathrm{t}$ is set by $S^\mathrm{b}_\nu$, which follows the Planck function. Or, instead, the \Hepsilon{} extinction dominates, and $S_\nu^\mathrm{t}$ follows $S^\mathrm{l}_{\nu_0}$, which is coupled to temperature through one of the terms in Eq. (\ref{eq:s_l}).

We address the formation of \Hepsilon{} with the forward synthesis of \Hepsilon{} from state-of-the-art radiative MHD simulations. We specify where we are looking in the solar atmosphere in terms of extinction and determine the dominant processes in the total source function. We evaluate the effects of Balmer line blanketing, HNEI, and 3D effects on the formation of \Hepsilon{}. In addition, we discuss the appearance of bright points and elongated dark fibrils apparent in the \Hepsilon{} images with example spectra shown in Fig.~\ref{fig:observational_profiles} and why and when we see \Hepsilon{} in emission against the \ion{Ca}{ii}~H background. We mainly focus on quiet Sun regions but will try to give some idea about the formation of \Hepsilon{} in more active regions.

\subsection{Balmer continuum}
\label{sec:Balmer continuum}

The Balmer continuum radiation is the main driver of the hydrogen rate system and the dominant source of hydrogen ionization in the solar chromosphere \citep{Carlsson2002}. Electrons recombine into high energy levels and cascade downwards, strongly affecting the transition rates and level populations of the hydrogen atom and, therefore, the \Hepsilon{} line. It is thus essential to properly model the Balmer continuum radiation field including line blends. RH 1.5D has the option to add line blends in the form of a Kurucz line list \footnote{For details see:  \url{http://kurucz.harvard.edu/linelists.html}}. The spectral lines from the Kurucz line list are calculated assuming LTE, including scattering. To speed up the computation we created a list of lines that affect the Balmer continuum radiation significantly. 

 \begin{figure}
    \centering
    \includegraphics[width=0.5\textwidth]{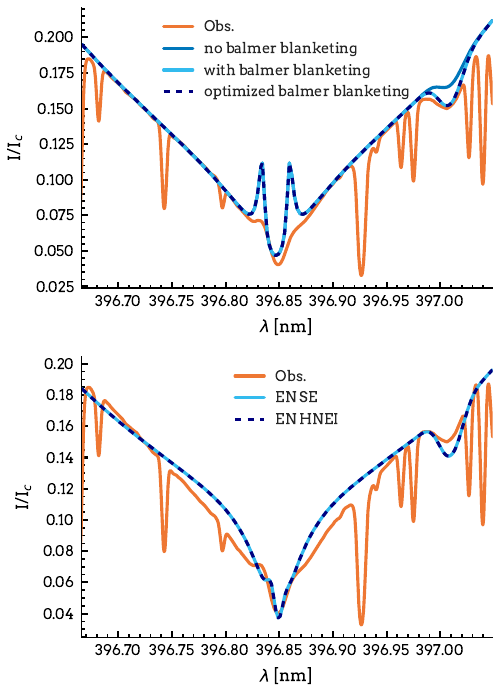}
    \caption{Synthetic \ion{Ca}{ii}~H and \Hepsilon{} spectral profiles compared against a FTS atlas observation. \emph{Top panel:} FTS atlas observations (orange) compared to the synthesised \ion{Ca}{ii}~H plus \Hepsilon{} profiles from the FALC model. Blue profile for the case with no Balmer continuum blanketing, light blue for Balmer blanketing with all lines included in the Kurucz line list, and dark blue for the optimized Kurucz line list including only lines affecting the Balmer continuum radiation significantly. \emph{Bottom panel:} Averaged Bifrost \ion{Ca}{ii}~H plus \Hepsilon{} profiles for the SE (light blue) and HNEI (dark blue) case against FTS atlas observations (orange).}
    \label{fig:Balmer_lineblanketing}
\end{figure}

Figure~\ref{fig:Balmer_lineblanketing} compares the synthesized \ion{Ca}{ii}~H plus \Hepsilon{} spectral region from the FALC model atmosphere against the FTS atlas normalized to the continuum. The inclusion of line blanketing in the Balmer continuum results in stronger \Hepsilon{} absorption closely matching the observations. The bottom panel compares the FTS atlas against the average line profile summed over the Bifrost simulation calculated with SE and HNEI. There is a strong effect of the Balmer radiation field strength on the \Hepsilon{} line profile, but this does not necessarily mean that the \Hepsilon{} source function is set by the Balmer radiation field. It only highlights that the hydrogen transitions (as \Hepsilon{}) are sensitive to changes in the Balmer radiation field. To show that the Balmer radiation field sets the \Hepsilon{} source function, it is necessary to show that the divergence between the upper and lower level of \Hepsilon{} with height is set by the Balmer radiation field. Further, we have to evaluate what sets the total source function in Eq. (\ref{eq:total_source_function}) and which term dominates the multi-level source function in Eq. (\ref{eq:s_l}) for \Hepsilon{}. 

 \begin{figure}
    \centering
    \includegraphics[width=0.5\textwidth]{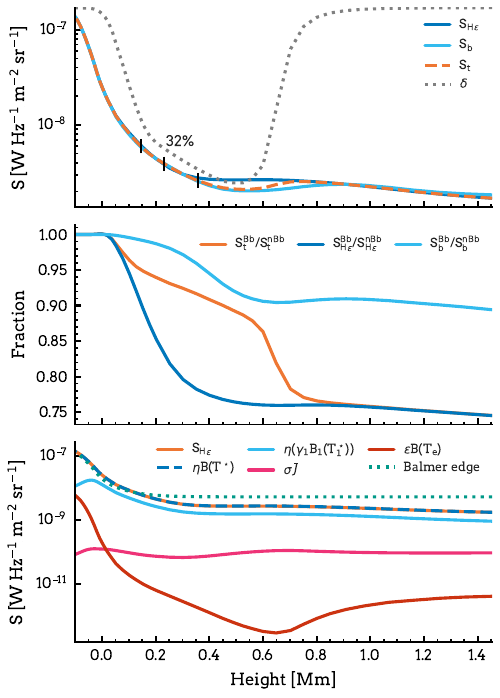}
    \caption{Total source function of \Hepsilon{}, effect of Balmer continuum line blanketing, and \Hepsilon{} multi-level source function for the FALC model. \emph{Upper panel:} Line ($S_{H\epsilon}$), background ($S_{b}$), and total source function ($S_{t}$), for the FALC model including Balmer continuum line blanketing. The dotted grey line shows the ratio $\delta$ between line extinction and total extinction (Eq. (\ref{eq:relative_opacity})). The three vertical ticks mark the total optical depth heights $\tau_\lambda=3,1,0.3$. The $32$\% label shows that at the $\tau_\lambda=1$ height $32$\% of the total source function consists of the \Hepsilon{} source function. \emph{Middle panel:} Fraction of the total, line, and background source function including Balmer blanketing and no Balmer blanketing. The change in the total source function with Balmer blanketing results from a change in the \Hepsilon{} source function, only making up $32$\% of the total source function. \emph{Bottom panel:} \Hepsilon{} source function split up in the multi-level source function terms (Eq. (\ref{eq:s_l}); scattering $\sigma \overline{J}$, thermal $\epsilon$B(T$_e$), and interlocking $\eta$B(T$_e$)). The \Hepsilon{} source function is dominated by interlocking and not by the Balmer edge (dotted green), with the hydrogen ground level as the dominant intermediate level ($\eta(\gamma$B$_1$(T$^\star_1$))). }
    \label{fig:Balmer_sf}
\end{figure}

Figure~\ref{fig:Balmer_sf} presents what sets the total source function (top panel), the influence of the Balmer line blanketing on the total source function (middle panel), and what sets the \Hepsilon{} source function for the FALC solar-atmosphere model. The total source function is dominated by the \ion{Ca}{ii}~H background source function throughout the lower solar atmosphere. At the formation height of \Hepsilon,{} only around $32\%$ of the photons are actually from the \Hepsilon{} transition. This stems from the fact that \ion{Ca}{ii}~H is a resonance line, whereas \Hepsilon{} lower level is an excited level with an energy jump of $10.2$~eV. The \ion{Ca}{ii}~H extinction will dominate the total extinction in the cooler parts of the atmosphere, with the dip in $\delta$ coinciding with the photospheric temperature minimum. This is an indication that the solar layers above reversed granulation are optically thin to \Hepsilon{} radiation.

\begin{figure*}
    \centering
    \includegraphics[width=1\textwidth]{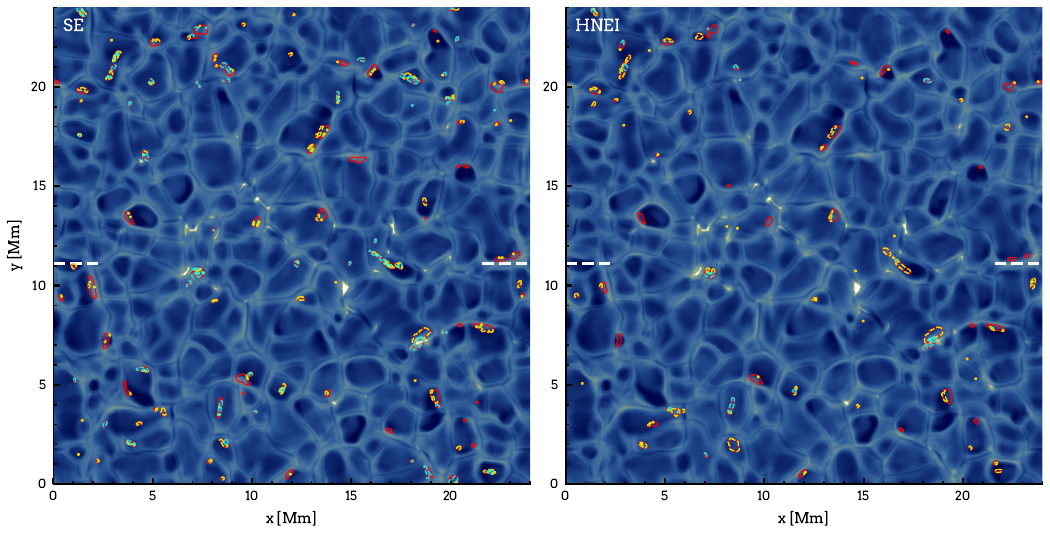}
    \caption{Synthetic \Hepsilon{} line core images at $397.1202$~nm. \emph{Left panel:} SE case. \emph{Right panel:} HNEI case. Contours mark regions of \Hepsilon{} emission at different column mass regions. \emph{Yellow:} column mass $\geq$ $2 \cdot 10^{-1}$~g cm$^{-2}$.  \emph{Red:} $2 \cdot 10^{-1}$~g cm$^{-1} >$ column mass $> 2 \cdot 10^{-2}$~g cm$^{-2}$.  \emph{Cyan:} column mass $\leq$ $2 \cdot 10^{-2}$~g cm$^{-2}$. \emph{Dashed lines:} position of the vertical slice shown in Fig.~\ref{fig:SE_HNEI_cut272_ion}.}
    \label{fig:EN_SE_HNEI}
\end{figure*}

The strong effect on the emergent \Hepsilon{} absorption stems from the Balmer line blanketing changing the total source function and not the extinction. The Balmer line blanketing has minor effects on the lower level populations of the \Hepsilon{} and \ion{Ca}{ii}~H transitions that make up the extinction. As a result, the atmospheric layers we are looking at in the FALC model do not change, but the number of photons available as described by the source function does change. The middle panel illustrates that the change in the total source function, resulting from the Balmer line blanketing is mainly due to a change in the \Hepsilon{} source function -- and not the background source function. Even when the total source function is dominated by the \ion{Ca}{ii}~H source function. An increase in the strength of the Balmer radiation field (no Balmer blanketing) will lead to an increase in the upper-level population of the \Hepsilon{} transition, therefore, increasing the source function but keeping the \Hepsilon{} extinction the same. However, this does not show that the \Hepsilon{} source function is set by the Balmer radiation field, only that the \Hepsilon{} source function is somehow affected by the Balmer radiation field. To illustrate which process dominates the \Hepsilon{} source function we have to evaluate the multi-level source function.

The bottom panel of Fig. \ref{fig:Balmer_sf} divides the \Hepsilon{} source function into the multi-level source function terms. The contributions from the scattering $\sigma \overline{J}$ and thermal $\epsilon B(T_\mathrm{e}$) part are orders of magnitude lower than the one from interlocking $\eta B(T_\mathrm{e}$). Interlocking dominates the \Hepsilon{} source function throughout the whole FALC model atmosphere. To highlight that the \Hepsilon{} source function is actually not following the Balmer radiation field we plotted the Balmer mean radiation field close to the Balmer edge in dotted green. The dominant interlocking term in Eq. (\ref{eq:bstar_seperate}) is $\gamma_1B_1(T^\star_1)$, with the ground level as an intermediate level. The Planck function $B_1(T^\star_1)$ is set by the ratio $P_{71}q_{12,7}/P_{21}q_{17,2}$, where the probability $q_{17,2}$ characterizes the \Hepsilon{} source function. The probability $q_{17,2}$ represents the sum of the transition probabilities, $p_{1i}$, Eq. (\ref{eq:indirect_seeq}) from the ground level into other hydrogen levels multiplied by the indirect transition probability to end up in the upper level of \Hepsilon{}. The transition probabilities are dominated by the radiative rates out of the ground level, not the collisional rates. 

In summary, in the FALC model, the total source function is dominated by the \ion{Ca}{ii}~H source function at the formation height of \Hepsilon. Our multi-level source function description suggests that the \Hepsilon{} source function is dominated by interlocking, with the ground level being the dominant intermediate level radiatively populating the upper level of \Hepsilon{} and not by the Balmer continuum radiation field as suggested by \citet{Ayres1975}.

\subsection[Effect of HNEI on Hepsilon]{Effect of HNEI on \Hepsilon{}}
\label{Effect of HNEI}

The solar chromosphere is pervaded by upwardly propagating magneto-acoustic shocks, making the effect of time-dependent hydrogen ionization important already above the solar photosphere. Non-equilibrium computations of hydrogen ionization strongly affect the temperature structure and ionization of the atmosphere due to the slow recombination rate at low-temperature intershock phases \citep{Leenaarts2007}. The ion population in turn is strongly coupled to the $n=2$ population of hydrogen, which sets the line extinction of \Hepsilon{} and therefore the formation height of Balmer series lines \citep{Carlsson2002}. In some cases, HNEI can  have strong effects depending on where we are looking at in the atmosphere and what we are observing, thereby directly influencing the \Hepsilon{} source function.

Figure \ref{fig:EN_SE_HNEI} shows synthesized \Hepsilon{} line core (at $397.1202$~nm) images from the Bifrost simulation (left panel for SE, and right panel for HNEI). The SE case uses the standard version of RH to solve the statistical equilibrium equation, which loses information about the time-dependent hydrogen ionization. Our HNEI implementation in RH uses the proton density given by the Bifrost simulation (see Sect. \ref{sec:stat_equil}). 

\begin{figure*}
    \centering
    \includegraphics[width=1\textwidth]{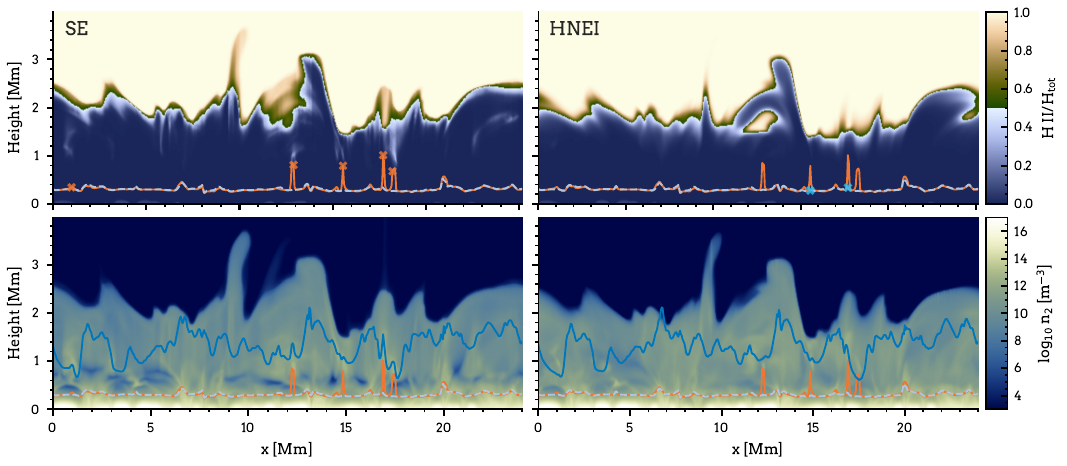}
    \caption{Hydrogen ionization (\emph{top}) and $n=2$ populations (\emph{bottom}) from a vertical slice through the simulation (indicated in Fig.~\ref{fig:EN_SE_HNEI}). \emph{Left column}: Results for statistical equilibrium (SE).  \emph{Right column}: Results for non-equilibrium hydrogen ionization (HNEI). The solid orange and dashed grey lines show the $\tau=1$ heights for both cases, and the orange crosses and light blue crosses mark regions where \Hepsilon{} is in emission. The dark blue line shows the relative variation of the \Hepsilon{} line core intensity.}
    \label{fig:SE_HNEI_cut272_ion}
\end{figure*}

The two images look almost identical. Both are mapping reversed granulation, which originates in the mid-photosphere. The synthetic images match the observed \Hepsilon{} images from Fig.~\ref{fig:observations_Hepsilon} quite well, which suggests a similar formation of \Hepsilon{} between simulation and observations. However, the effect of HNEI seems negligible for the observed reversed granulation pattern, which in the simulation is mainly made up of weak \Hepsilon{} absorption lines. To evaluate the effect of HNEI on the \Hepsilon{} source function and extinction, we look at a vertical cut through the Bifrost simulation, denoted by two dashed lines in Fig. \ref{fig:EN_SE_HNEI}.  

\begin{figure}
    \centering
    \includegraphics[width=0.5\textwidth]{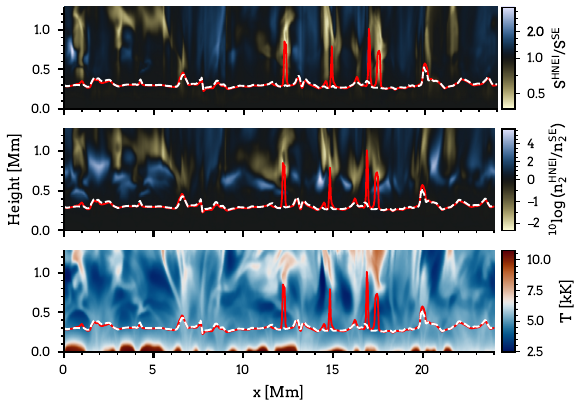}
    \caption{Departures from statistical equilibrium from a vertical slice through the Bifrost simulation. \emph{Top panel:} Departure of the total source function from SE. \emph{Middle panel:} Departure of the hydrogen $n=2$ populations from SE. \emph{Bottom panel:} Temperature structure of the Bifrost simulation. The vertical slice is taken at the position marked with white dashed lines in Fig.~\ref{fig:EN_SE_HNEI}. \emph{Red solid:} \Hepsilon{} $\tau=1$ height for SE. \emph{White dashed:} \Hepsilon{} $\tau=1$ height for HNEI.}
    \label{fig:SE_HNEI_cut272_source}
\end{figure}

Figure \ref{fig:SE_HNEI_cut272_ion} compares the hydrogen ionization structure and $n=2$ level population for the cases of SE and HNEI. The HNEI case shows a different and smoother ionization structure through the atmosphere compared to the SE case. This is also reflected in the $n=2$ populations -- especially in the lower part of the atmosphere (below $1$~Mm), where the dark voids vanish under HNEI. The reason for the smoother $n=2$ populations is that under HNEI the $n=2$ populations follow the hydrogen ionization \citep{Leenaarts2007}, whereas for SE, the $n=2$ population follows the temperature (compare temperature panel in Fig. \ref{fig:SE_HNEI_cut272_source}). This temperature sensitivity below $1$~Mm stems from the fact that Lyman-$\alpha$ is in radiative balance $n_1/n_2=R_{21}/R_{12}$ with $S \approx J \approx B$. Therefore, the $n=2$ hydrogen population follows LTE: Saha-Boltzmann partitioning for the predominately neutral lower solar atmosphere, with high Lyman-$\alpha$ extinction \citep{Vernazza1981, Rutten2016, Rutten2017}.

The $n=2$ level population increases strongly for hot pockets in the lower solar atmosphere in SE. This raises the $\tau=1$ height, represented by the orange line in Fig. \ref{fig:SE_HNEI_cut272_ion}. These elevated $\tau=1$ locations are connected to \Hepsilon{} emission indicated by crosses. Under HNEI the \Hepsilon{} extinction does not ``feel'' this temperature increases as much as in SE. This not only has a strong influence on the extinction but also on the source function itself. The influence of HNEI on the $n=2$ populations and total source function in terms of departure from SE can be seen in Fig. \ref{fig:SE_HNEI_cut272_source}. We show the departures from SE for the lower solar atmosphere in the formation region of \Hepsilon{}. The $\tau=1$ heights for \Hepsilon{}, where most of the radiation escapes are similar between SE and HNEI, which illustrates that the increase in the $n=2$ populations under HNEI is not enough to change where we are looking at in the simulation. Further, the total source functions at the $\tau=1$ height do not change much between SE and HNEI. Therefore, time-dependent hydrogen ionization effects are negligible for the \Hepsilon{} extinction and the total source function where \Hepsilon{} absorption is formed. However, this is not the case for \Hepsilon{} emission.

\Hepsilon{} emission is co-located with enhanced temperatures in the lower solar atmosphere, as illustrated in the lower panel of Fig.~\ref{fig:SE_HNEI_cut272_source} (\Hepsilon{} emission is marked with crosses in Fig. \ref{fig:SE_HNEI_cut272_ion}). This leads to two orders of magnitude higher $n=2$ populations at these heating events, shifting up the $\tau=1$ formation heights. At these particular atmospheric heights, the total source function under SE is significantly higher (up to a factor of 2) compared to HNEI, resulting in stronger \Hepsilon{} emission. Only two out of five emission locations (crosses in Fig. \ref{fig:SE_HNEI_cut272_ion}) are still in emission in HNEI. Even though the $\tau=1$ height is lower in HNEI, for these two locations in emission, under SE, the \Hepsilon{} emission profile has a significant contribution from these higher-lying layers, where the total source function is increased.

HNEI has a sizeable effect on the $n=2$ populations throughout the whole atmosphere, from the lower atmosphere up to the transition region. It changes not only the extinction but also the \Hepsilon{} source function. This effect can be seen in Fig.~\ref{fig:SE_HNEI_cut272_source} (top panel). Similarly, we find that HNEI also affects the line source functions of other lines in the Balmer series (not shown in the Figures).
Depending on where the lines are formed, the difference between the emergent Balmer series line profiles in SE or HNEI can be significant. Another important side effect of HNEI is time-dependent ionization, where the hydrogen populations can have a memory of a previous atmospheric state, decoupling the observed line profiles from the instantaneous conditions in the atmosphere. This has implications for the interpretation of \Halpha{} spectroheliograms, as they contain information about previous events, especially heating events that increase the \Halpha{} extinction \citep{Rutten2016}.

\begin{figure}
    \centering
    \includegraphics[width=0.5\textwidth]{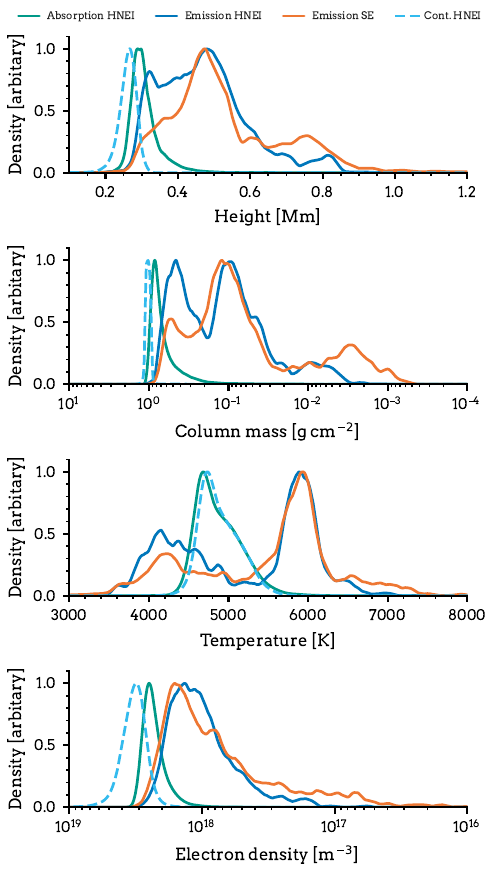}
    \caption{Distributions of atmospheric quantities at the \Hepsilon{} $\tau=1$ height.  Panels show the kernel density estimation as a function of height, column mass, temperature, and electron density at the $\tau=1$ height for the \Hepsilon{} rest wavelength. The distributions are separated by the type of \Hepsilon{ } profile, as indicated in the legend: absorption in HNEI,  emission in HNEI, emission in SE, and the \ion{Ca}{ii}~H background continuum at \Hepsilon{} rest wavelength in HNEI.}
    \label{fig:kernel_density}
\end{figure}

The strong effect of HNEI on \Hepsilon{} emission can be seen in Fig. \ref{fig:EN_SE_HNEI}, where the over-plotted contours mark regions where \Hepsilon{} is in emission. The different colored contours group the column masses at $\tau=1$ height into three different groups related to the column mass panel of Fig. \ref{fig:kernel_density}. The yellow contours highlight regions where the column mass at $\tau=1$ is greater than or equal to $2 \cdot 10^{-2}$~g cm$^{-1}$, red regions greater than $2 \cdot 10^{-1}$~g cm$^{-2}$ and less than $2 \cdot 10^{-2}$~g cm$^{-2}$, and the cyan color less than or equal to $2 \cdot 10^{-2}$~g cm$^{-2}$. Figure \ref{fig:kernel_density} shows a kernel density estimation \citep{Rosenblatt1956} for the distributions of formation heights, column masses, temperatures, and electron densities at the $\tau = 1$ heights for \Hepsilon{} emission and  absorption lines for the Bifrost simulation. \Hepsilon{} in absorption is formed just above the \ion{Ca}{ii}~H background, whereas \Hepsilon{} emission is formed across a large range of column masses. This range is connected to enhanced temperatures in the lower solar atmosphere, which shifts the formation height upwards under SE  and is connected to increased ionization in HNEI. Generally, HNEI reduces the amount of \Hepsilon{} emission formed higher up in the atmosphere.

In summary, \Hepsilon{} synthetic images are similar to solar observations, showing mainly reversed granulation. Why we observe reversed granulation in the \Hepsilon{} line core, we will address in the Sect. \ref{sec:Hepsilon absorption}. Furthermore, we have to explain the small bright points seen in the \Hepsilon{} line core images; HNEI has little effect on the reversed granulation pattern but influences the \Hepsilon{} emission. As \Hepsilon{} emission can form relatively high up in the lower solar atmosphere, we have to check for probable 3D radiative transfer effects (Sect. \ref{sec:3D effects}).

\subsection[Hepsilon absorption]{\Hepsilon{} absorption}
\label{sec:Hepsilon absorption}

In this section, we discuss the formation of \Hepsilon{} absorption profiles. We analyze why we observe reversed granulation in weak \Hepsilon{} absorption profiles and why strong \Hepsilon{} absorption lines are linked to magnetic elements.

\begin{figure}
    \centering
    \includegraphics[width=0.5\textwidth]{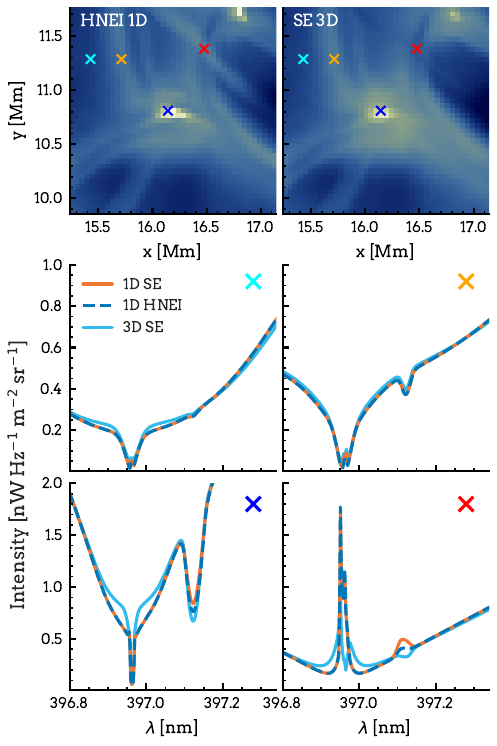}
    \caption{Synthetic \ion{Ca}{ii}~H and \Hepsilon{} images and spectral profiles for reversed granulation, magnetic elements, and a dark fibrilar structure. \emph{Topmost row:} Synthesized \Hepsilon{} line core images from a cutout of the Bifrost simulation calculated in 1D (\emph{left}) and 3D (\emph{right}). \emph{Bottom two rows:} \ion{Ca}{ii}~H plus \Hepsilon{} spectral profiles from locations marked with coloured crosses in the \Hepsilon{} line core images. \emph{Middle-left:} Granular spectral profile. \emph{Middle-right:} Intergranular spectral profile. \emph{Bottom-left:} Spectral profile located at a magnetic field concentration. \emph{Bottom-right:} Location where \Hepsilon{} emission in 1D turns to absorption in 3D.}
    \label{fig:absorption_granule_magnetic}
\end{figure}

Figure \ref{fig:absorption_granule_magnetic} shows a synthetic \Hepsilon{} line core image from a small region from the Bifrost simulation. The region covers a magnetic element and parts of the reversed granulation pattern. To check for 3D effects we synthesized the same region in 3D, but assuming SE and treating \ion{Ca}{ii}~H in CRD. We adopted these approximations because it was difficult to get RH to converge under HNEI and PRD in regions with strong horizontal inhomogeneities, such as magnetic elements. Representative profiles are shown in Fig. \ref{fig:absorption_granule_magnetic} including one particularly interesting structure, which goes from \Hepsilon{} emission in 1.5D to absorption in 3D (this structure crosses the red cross at an approximately 45$^\circ$ angle). We address the effects of 3D radiative transfer on \Hepsilon{} emission in Sect.~\ref{sec:3D effects}.

\begin{figure*}
    \centering
    \includegraphics[width=1\textwidth]{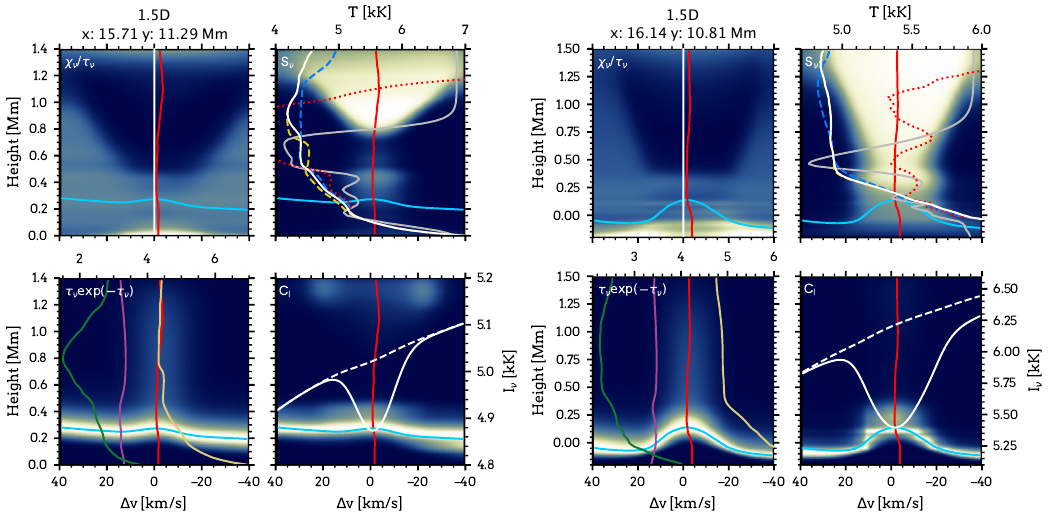}
    \caption{\Hepsilon{} four-panel diagrams for an intergranular lane (\emph{left}) and a magnetic bright point (\emph{right}). The locations of each point in the simulation are marked with orange and blue crosses in Fig.~\ref{fig:absorption_granule_magnetic}. Each diagram is organized as follows. \emph{Top left panels:} Relative $\chi_\nu/\tau_\nu$; sensitive to velocity gradients. \emph{Top right panels:} $S_\nu$: Relative source function. \emph{Solid white:} Total source function at rest wavelength. \emph{Dashed yellow:} \Hepsilon{} source function. \emph{Dashed blue:} Background source function. Source functions are displayed in brightness temperature units with the scale at the top.  \emph{Solid grey:} Ratio of \Hepsilon{} to total extinction (if \Hepsilon{} extinction dominates the grey curve is on the left side of the plot; if \ion{Ca}{ii}~H extinction dominates on the right). \emph{Dotted red:} Atmospheric temperature profile. \emph{Bottom-left panels:} Relative $\tau_\nu e^{(-\tau_\nu)}$; indicates where most of the contribution to relative line intensity comes from. \emph{Solid green:} $\epsilon B_\nu (T_e)$. \emph{Solid violet:} $\sigma J_\nu$. \emph{Solid gold:} $\eta B_\nu (T^\star)$. The three terms are displayed in brightness temperature units with the scale at the top. \emph{Bottom-right panels:} $C_I$; contribution to relative line depression. \emph{Solid white:} \ion{Ca}{ii}~H plus \Hepsilon{} line profile. \emph{Dashed white:} \ion{Ca}{ii}~H background line profile. \emph{Solid red:} Atmospheric vertical velocity profile. \emph{Solid light blue:} $\tau=1$ height. } 
    \label{fig:4PFD_absorption_magnetic_element}
\end{figure*}

In general, we observe very weak \Hepsilon{} absorption lines inside granules that get stronger the closer we get to the granular edge. 3D effects are negligible on the \Hepsilon{} line profiles (see middle panels of Fig.~\ref{fig:absorption_granule_magnetic}). To discuss the formation of weak \Hepsilon{} absorption lines we created relative four-panel formation diagrams inspired by \cite{Carlsson1997}. Figure \ref{fig:4PFD_absorption_magnetic_element} displays such diagrams, which we henceforth refer to as ``four-panel diagrams.'' The four-panel diagrams decompose the contribution to relative line depression or emission into three components:

\begin{equation}\label{eq:R4PFD_decompose}
    C_\mathrm{I} = \frac{\mathrm{d}I_\mathrm{R}(\nu,z)}{\mathrm{d}z} = \frac{\chi^\mathrm{R}_\nu(z)}{\tau^\mathrm{R}_\nu(z)} \cdot \tau^\mathrm{R}_\nu(z) \, \mathrm{exp}(-\tau^\mathrm{R}_\nu(z)) \cdot S^\mathrm{R}_\nu(z),
\end{equation} 
based on Eq. (\ref{eq:relative cf}). 
The first component $\chi^\mathrm{R}_\nu(z)/\tau^\mathrm{R}_\nu(z)$ is the ratio between relative extinction and relative optical depth. This term has large values where there are many absorbing or emitting particles (line and background; $\chi^\mathrm{l}_\nu + \chi^\mathrm{b}_\nu$) at low relative optical depth. This component is sensitive to line-of-sight velocity gradients that move plasma, so that they effectively increase extinction at varying Doppler offsets at wavelength regions with low optical depths.

The second component $\tau^\mathrm{R}_\nu(z) \, \mathrm{exp}(-\tau^\mathrm{R}_\nu(z))$ has the the largest values around the $\tau=1$ height. In this panel, we over-plotted the three components of the multi-level source function (Eq. \ref{eq:s_l}) for \Hepsilon{}. This helps to determine the dominant line-formation process. 

The third component $S^\mathrm{R}_\nu(z)$ describes the relative source function. The relative source function can have positive or negative values. This is a major difference from standard four-panel formation diagrams, where the source function panel only has positive values. Here, positive values signify that the line source function, $S^\mathrm{l}_\nu$, is lower than the background intensity, $I^\mathrm{b}_\nu$, resulting in a contribution to relative line depression at a particular height in the atmosphere. Negative values signify a relative contribution to line emission. However, the relative source function does not specify what sets the total source function at the formation height of \Hepsilon{}. Based on the upper panel of Fig \ref{fig:Balmer_sf}, we included the different components expressing the total source function at the \Hepsilon{} line core wavelength: the total source function, the \Hepsilon{} source function, the background source function, and additionally included the ratio of \Hepsilon{} extinction to total extinction.
The last panel shows $C_I$, which is the contribution to the relative line depression or emission. We note that $C_I$ is positive in regions that contribute to line absorption and negative in regions that contribute to line emission. 

 First, we focus on the left four-panel diagram in Fig.~\ref{fig:4PFD_absorption_magnetic_element}, which illustrates the formation of a relatively weak \Hepsilon{} absorption line in the intergranular lane (orange cross in Fig.~\ref{fig:absorption_granule_magnetic}). Most contribution to relative line depression is formed just above the \ion{Ca}{ii}~H background intensity. This highlights that the added \Hepsilon{} extinction on top of the \ion{Ca}{ii}~H extinction is very small, making the atmosphere nearly transparent to \Hepsilon{} radiation. Further, we have to evaluate what sets the total source function at the line formation region specified by the relative extinction over-plotted in the relative source function panel. The grey curve points out that more than $50\%$ of line photons are actually coming from the \ion{Ca}{ii}~H transition. The total source function is more tightly coupled to the \ion{Ca}{ii}~H source function than to the \Hepsilon{} source function. The \ion{Ca}{ii}~H source function is still strongly coupled to LTE, which means that in this case more than $50\%$ of photons at the \Hepsilon{} wavelength are formed under LTE conditions. The optical depth $\tau_\lambda$ of the atmosphere above the height where the \ion{Ca}{ii}~H background is formed is just below unity ($\tau\approx 0.78$ for the intergranular location), thus giving the appearance of reversed granulation in the \Hepsilon{} line core, which happens for most for quiet Sun locations.

The closer we get to the granular center, the weaker the \Hepsilon{} absorption gets (see the middle row in Fig. \ref{fig:absorption_granule_magnetic}). This stems from the fact that the mid-photosphere is cooler at the center of granules where reversed granulation is formed. As the \Hepsilon{} extinction is strongly coupled to Saha-Boltzmann due to Lyman-$\alpha$ thermalization, the \Hepsilon{} contribution to the total extinction becomes smaller, essentially removing the \Hepsilon{} line depression. The decrease in \Hepsilon{} extinction will couple the total source function even closer to the \ion{Ca}{ii}~H source function ($\approx$80\%).

The change in \Hepsilon{} absorption strength from the granular center to the granular edge is therefore dependent on the temperature structure and not so much on the strength of the Balmer radiation. The Balmer radiation will indeed increase the \Hepsilon{} source function in both SE and HNEI. In SE, a stronger Balmer radiation field will not change the \Hepsilon{} extinction ($n=2$ population), but increase the $n=7$ populations, thereby increasing the \Hepsilon{} source function. In HNEI, the behavior is different: the \Hepsilon{} extinction and $n=7$ populations  decrease for a stronger Balmer radiation field but still result in an increased \Hepsilon{} source function as the \Hepsilon{} extinction decreases more rapidly with height than the $n=7$ populations. However, the \Hepsilon{} source function for weak \Hepsilon{} absorption is dominated by interlocking illustrated by the gold solid line in the $\tau^\mathrm{R}_\nu(z) \, \mathrm{exp}(-\tau^\mathrm{R}_\nu(z))$ panel. Similarly, as in Sect. \ref{sec:Balmer continuum}, our multi-level source function description suggests the interlocking term is dominated by the Lyman series with indirect transitions into the upper level \Hepsilon{} through the ground level -- and not by the Balmer radiation field. The Balmer radiation field will affect the mean radiation fields in the Lyman series, but not the \Hepsilon{} source function directly.

We now move to the formation of strong \Hepsilon{} absorption lines. As a representative example, we consider the bottom left line profile in Fig. \ref{fig:absorption_granule_magnetic} (blue cross), and show its corresponding four-panel diagram in Fig. \ref{fig:4PFD_absorption_magnetic_element} (right panels). This strong \Hepsilon{} absorption line is connected to a magnetic field concentration. Due to the low gas density inside magnetic elements, \Hepsilon{} forms at lower heights, therefore, it is associated with higher temperatures. As the \ion{Ca}{ii}~H wing intensity is well approximated by LTE and follows the temperature, the intensity above the magnetic elements will be significantly increased compared to that in reversed granulation. The reason is similar to why magnetic elements appear bright in the wings of \Halpha{} \citep{Leenaarts2006}. However, \Hepsilon{} is formed on top of the \ion{Ca}{ii}~H wing. The added \Hepsilon{} extinction shifts the formation height upwards making magnetic elements optically thick to \Hepsilon{} radiation. The total source function strongly decreases with height, being mainly set by the \Hepsilon{} source function due to the increased \Hepsilon{} extinction above magnetic elements. This mapping of higher layers with a lower source function compared to the \ion{Ca}{ii}~H background leads to strong \Hepsilon{} absorption lines. Therefore, \Hepsilon{} can be used to identify and track weak magnetic elements.

Our multi-level source function description also suggests that the \Hepsilon{} source function is dominated by interlocking. The dominant interlocking process is an indirect transition loop through the ground level for strong \Hepsilon{} absorption profiles connected to magnetic field concentrations, the same process as found for weak \Hepsilon{} absorption profiles.

The fourth profile, marked with a red cross in Fig. \ref{fig:absorption_granule_magnetic} is a good example that some \Hepsilon{} profiles can suffer significant 3D effects. In this particular case, the bright elongated structure seen in 1D will turn into a faint fibrilar structure if 3D radiative transfer effects are taken into account. In the next section, we address 3D effects on \Hepsilon{} emission regions, where the formation height is increased compared to the usual reversed granulation.

\subsection[3D effects on Hepsilon emission]{3D effects on \Hepsilon{} emission}
\label{sec:3D effects}

The Balmer series lines, for example \Halpha{}, are strongly affected by scattering. The characteristic fibrilar chromosphere seen in \Halpha{} can only be reproduced if the radiation field is treated in 3D geometry \citep{Leenaarts2012}. However, as shown in Sect.~\ref{sec:Balmer continuum}, the \Hepsilon{} source function is not dominated by two-level scattering but instead by interlocking through the Lyman series. The Lyman series lines are the strongest scattering lines in the solar spectrum, having the lowest collisional destruction probability per line photon extinction. Hence, we evaluated 3D Lyman scattering effects on the emergent \Hepsilon{} line profiles formed higher up in the atmosphere (at lower column masses than $2 \cdot 10^{-2}$~g~cm$^{-2}$).

\begin{figure}
    \centering
    \includegraphics[width=0.5\textwidth]{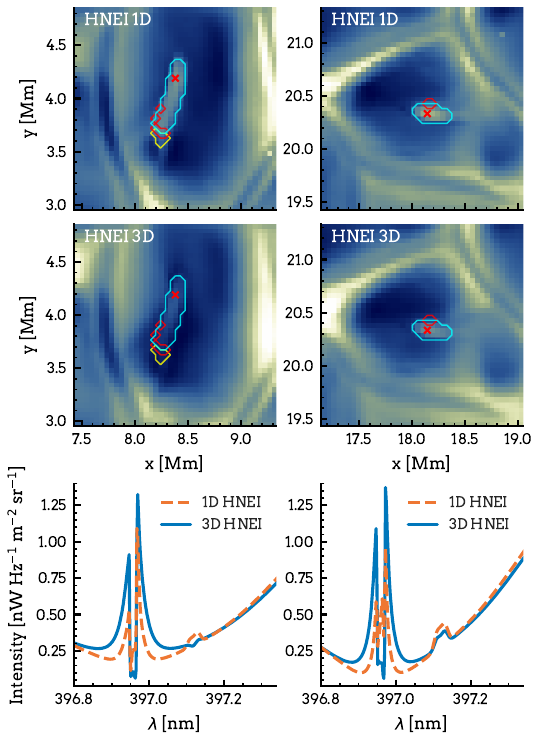}
    \caption{Synthetic \ion{Ca}{ii}~H and \Hepsilon{} images and spectral profiles for \Hepsilon{} emission formed at low mass densities. \emph{Top two rows:} Synthesized \Hepsilon{} line core images from a cutout of the Bifrost simulation calculated in 1D (\emph{top row}) and 3D (\emph{middle row}). Contours mark column mass regions, same as in Fig.~\ref{fig:EN_SE_HNEI}. \emph{Bottom row:} \ion{Ca}{ii}~H plus \Hepsilon{} spectral profiles from locations marked with red crosses in the \Hepsilon{} line core images.}
    \label{fig:HNEI_1D_3D_high_coldens}
\end{figure}

\begin{figure*}
    \centering
    \includegraphics[width=1\textwidth]{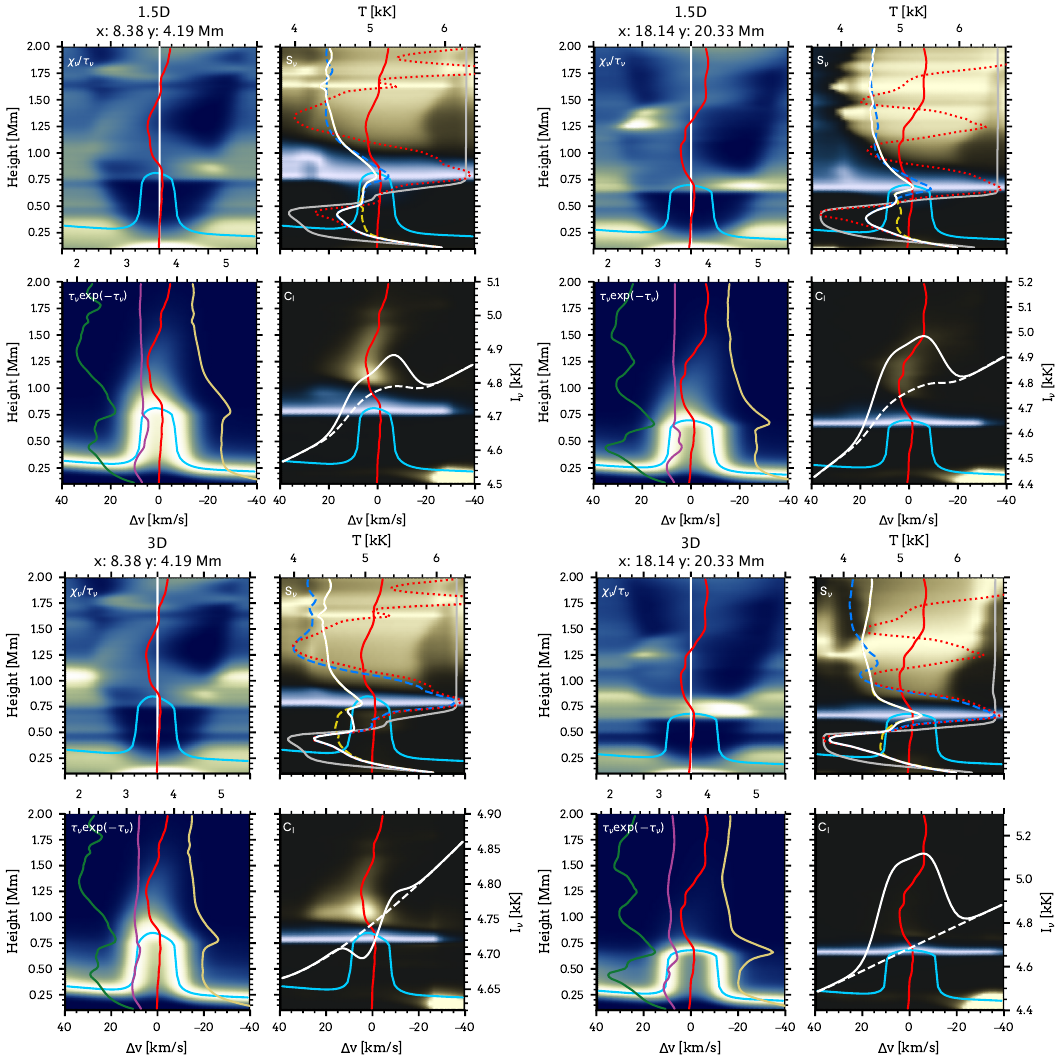}
    \caption{\Hepsilon{} four-panel diagrams for the locations marked with crosses in Fig. \ref{fig:HNEI_1D_3D_high_coldens}. \emph{Top row:} \Hepsilon{} synthesized in 1.5D. \emph{Bottom row:} Same location synthesized in 3D. \emph{Top left panels:} Relative $\chi_\nu/\tau_\nu$; sensitive to velocity gradients. \emph{Top right panels:} $S_\nu$; relative source function. Yellow colormap for negative source function values. Light blue colormap for positive source function values. \emph{Solid white:} Total source function at rest wavelength. \emph{Dashed yellow:} \Hepsilon{} source function. \emph{Dashed blue:} Background source function. Source functions are displayed in brightness temperature units with the scale at the top. \emph{Solid grey:} Ratio of \Hepsilon{} to total extinction (if \Hepsilon{} extinction dominates the grey curve is on the left side of the plot; if \ion{Ca}{ii}~H extinction dominates on the right). \emph{Dotted red:} Atmospheric temperature profile. \emph{Bottom left panels:} Relative $\tau_\nu e^{(-\tau_\nu)}$; indicates where most of the contribution to relative line intensity comes from. \emph{Solid green:} $\epsilon B_\nu (T_e)$. \emph{Solid violet:} $\sigma J_\nu$. \emph{Solid gold:} $\eta B_\nu (T^\star)$. The three terms are displayed in brightness temperature units with the scale at the top. \emph{Bottom right panels:} $C_I$; contribution to relative line depression/emission. Yellow colormap contribution to line depression. Light blue colormap contribution to line emission. \emph{Solid white:} \ion{Ca}{ii}~H plus \Hepsilon{} line profile. \emph{Dashed white:} \ion{Ca}{ii}~H background line profile. \emph{Solid red:} Atmospheric upward velocity profile. \emph{Solid light blue:} $\tau=1$ height.}
    \label{fig:4PFD_x176_x381}
\end{figure*}

To investigate 3D effects on the emergent \Hepsilon{} profiles, we chose small regions of interest from the Bifrost simulation. From these regions, we synthesized \Hepsilon{} with the assumption of HNEI in 1D and 3D geometry. We carried out spectral synthesis in 3D only in these selected regions to avoid the large computational costs of a full 3D synthesis, and because there are only a few regions where \Hepsilon{} is formed higher up in the atmosphere. Furthermore, we synthesized \ion{Ca}{ii}~H in CRD instead of PRD in 3D geometry as the running with PRD led to numerical instabilities and the problem often failed to converge. Because we assume CRD in 3D, the intensity at the \ion{Ca}{ii}~H inner wings will deviate between 3D and 1.5D.

First, we discuss the case of low column mass, where 3D effects become important at certain locations. Figure \ref{fig:HNEI_1D_3D_high_coldens} shows two regions of interest from the Bifrost simulation, where \Hepsilon{} is in emission at low column mass. The left column shows a region with strong 3D effects, while the right column shows a region with negligible 3D effects. 
We show the corresponding four-panel diagrams in Fig. \ref{fig:4PFD_x176_x381}.

The left column of Fig. \ref{fig:4PFD_x176_x381} illustrates the formation of \Hepsilon{} with strong 3D effects. In the 3D case, $C_\mathrm{I}$ outlines significantly more contribution to line depression than for the 1.5D case on top of the relative line emission region. 
The answer to the question of why we have more contribution to line depression in 3D is twofold. First the \Hepsilon{} extinction $\chi^\mathrm{l}_\nu$ plays an important role (see. Eq. (\ref{eq:relative_contr})). The $n=2$ populations are in radiative equilibrium with Lyman-$\alpha$ as $n_2 = n_1 R_{12}/R_{21}$ is set by the Lyman-$\alpha$ mean radiation field and the exponential density decrease of the atmosphere. The Lyman-$\alpha$ mean radiation field in 3D is slightly increased compared to 1.5D. Therefore, the \Hepsilon{} extinction is mapping more of the positive relative source function giving rise to relative line depression. However, the increase in \Hepsilon{} extinction is not the main reason for the enhanced contribution to the relative line depression in 3D highlighted by similar $\tau_\nu \mathrm{exp}(-\tau_\nu)$ panels. 

The main reason why we see enhanced relative line depression in 3D is that the \Hepsilon{} line source function is significantly lower than the 1.5D case above a certain atmospheric height. This decrease in $S^\mathrm{l}_\nu$ in 3D causes a negative relative line source function at lower heights compared to 1.5D (illustrated in the top right $S_\nu$ panels in Fig. \ref{fig:4PFD_x176_x381}). The relative source function becomes negative when the \Hepsilon{} source function $S^\mathrm{l}_\nu$ is smaller then the background \ion{Ca}{ii}~H wing intensity, specified by $1 - S^\mathrm{l}_\nu/I^\mathrm{b}_\nu$. The background intensity $I^\mathrm{b}_\nu$ flattens out with height (around $\approx 0.5$~Mm), having lower values in 3D. Our multi-level source function description suggests that the \Hepsilon{} source function is dominated by interlocking, set by the Lyman series (same process as for \Hepsilon{} absorption), or more precisely, a linear combination of the transition probability, $p_{1i}$, multiplied by an indirect transition probability, $q_{i7,2}$, due to a Lyman series transition. The transition probabilities are dominated by the radiative rates. The net rate out of the ground level is set by the Lyman-$\alpha$ radiative rate which is orders of magnitude higher than other Lyman series radiative rates. Therefore, the transition probabilities are given by $p_{1i} = R_{1i}/R_{12}$ and reflect the ratio between the Lyman series and the Lyman-$\alpha$ mean radiation field. The main difference between 1.5D and 3D is that this ratio decreases more strongly with height in 3D than in 1.5D. This puts the \Hepsilon{} source function, $S^\mathrm{l}_\nu$, below the \ion{Ca}{ii}~H background intensity, $I^\mathrm{b}_\nu$, at lower heights giving rise to a negative source function creating the strong relative absorption contribution in 3D.

The negligible 3D effects on \Hepsilon{} for the right column of Fig.~\ref{fig:4PFD_x176_x381} stem from the fact that the temperature increase is situated lower in the atmosphere than for the left column. This temperature rise at lower heights leads to a coupling of the Lyman-$\alpha$ mean radiation field with the Planck function to heights where the temperature is decreasing again, above the temperature peak. Whereas for the case of strong 3D effects, the Lyman-$\alpha$ mean radiation decouples from the Planck function at the temperature peak in the atmosphere. In both cases, the Lyman-$\alpha$ mean radiation decouples approximately at the same height from the Planck function.  As the $n=2$ level population follows the Lyman-$\alpha$ mean radiation via $n_2 = n_1 R_{12}/R_{21}$, the \Hepsilon{} line extinction $\chi^\mathrm{l}_\nu$ decreases after the temperature rise. This strong decrease in line extinction, as illustrated in the $\tau_\nu \mathrm{exp}(-\tau_\nu)$ panel, is the main reason why we do not see a strong contribution to relative line depression above the emission region. The \Hepsilon{} emission formed at greater column masses than $2 \cdot 10^{-2}$~g cm$^{-2}$ (yellow and red contours in Fig.~\ref{fig:EN_SE_HNEI}) show negligible 3D effects.

In summary: if a temperature increase in the lower solar atmosphere extends too high up in the atmosphere and is not confined in the lower solar atmosphere (for the Bifrost simulation this would be approximately a height of $0.75$~Mm), the \Hepsilon{} line will suffer from 3D effects, showing a \Hepsilon{} absorption line instead of a \Hepsilon{} emission line in 1.5D, given the density stratification of the Bifrost atmosphere.

This explains the strong change in the appearance of the bright fibrilar structure in 1.5D in Fig. \ref{fig:absorption_granule_magnetic}, compared with the dark fibrilar structure seen in 3D. Figure \ref{fig:SE_HNEI_cut272_source} show a temperature cut through this fibrilar structure (white dashed lines in Fig. \ref{fig:EN_SE_HNEI}). The strong temperature enhancement throughout the lower atmosphere connected to this structure leads to an increased \Hepsilon{} line extinction via $n_2 = n_1 R_{12}/R_{21}$. The 3D effects lead to a lower \Hepsilon{} line source function than the \ion{Ca}{ii}~H background intensity $I^\mathrm{b}_\nu$ and give a strong contribution to relative line depression above the emission region compared to the 1.5D case. As a result, the 1.5D \Hepsilon{} emission lines at the fibrilar structure become absorption lines in 3D.

\section{Discussion}
\label{sec:Discussion}

\subsection{The \Hepsilon{} source function}

Numerical experiments such as the Bifrost simulation represent a solar-like model atmosphere with properties close to the real solar atmosphere. We focus our discussion on the similarities between synthetic and observed \Hepsilon{} profiles. The synthetic \Hepsilon{} profiles combined with the Bifrost simulation can help us understand the basic formation processes in the solar atmosphere that lead to features observed in the quiet Sun, such as reversed granulation, bright points, \Hepsilon{} emission, and dark fibrilar structures.

Several decades ago, \citet{Thomas1957} suggested dividing the resonance lines and strong lines such as \Halpha{} into collisionally or photoelectrically dominated. For \Halpha{}, this division meant that the source function is set by collisions, the second term in Eq. (\ref{eq:s_l}), or by the third term (photoelectric) in Eq. (\ref{eq:s_l}). The third term describes the interlocking contribution to the line source function and, following \citet{Thomas1957}, it states that the indirect path that dominates the \Halpha{} source function is photoionization from $n=2$ and a recombination cascade into $n=3$ (thus, the description of it as photoelectric). However, \citet{Rutten2012} demonstrated that the \Halpha{} source function is dominated by scattering, described by the first term in Eq. (\ref{eq:s_l}), with strong contributions from chromospheric backscattering. However, our main question concerns which process dominates the source function of higher order Balmer series lines, such as \Hepsilon.

\citet{Ayres1975} first modeled the formation of \Hepsilon{} based on 1D numerical radiative transfer calculations for two model atmospheres, one representing the Sun (\Hepsilon{} in absorption) and one Arcturus (\Hepsilon{} in emission). They concluded that the Balmer continuum radiation dominates the \Hepsilon{} source function via photoionization, the same mechanism proposed by \citet{Thomas1957} for the \Halpha{} source function. However, our \Hepsilon{} forward modeling from the FALC and Bifrost model atmosphere suggests otherwise. Our multi-level source function description suggests that the \Hepsilon{} source function is dominated by interlocking via the ground level of hydrogen due to the Lyman series. Not the Balmer continuum radiation field.

Our synthetic \Hepsilon{} line core images (Fig. \ref{fig:EN_SE_HNEI}) reproduce the observed \Hepsilon{} reversed granulation pattern (Fig. \ref{fig:observations_Hepsilon}) quite well. Both show the characteristic dark granules and bright intergranular lanes suggesting that the Bifrost simulation has the basic physical properties representing a quiet Sun lower atmosphere. The reversed granulation pattern consists of weak \Hepsilon{} absorption lines. The weakest \Hepsilon{} absorption profiles can be found inside granules and the closer we get to the intergranular lanes, the stronger the \Hepsilon{} absorption becomes. We illustrated that the total source function for weak \Hepsilon{} absorption lines is dominated by the background \ion{Ca}{ii}~H source function and not the \Hepsilon{} source function. The temperature structure of the reversed granulation sets the extinction and total source function and therefore the behavior of weak \Hepsilon{} absorption lines.

We tested the influence of HNEI on the emergent \Hepsilon{} line profiles against SE. The influence is negligible at the reversed granulation pattern, which is represented by weak \Hepsilon{} absorption lines, formed deep in the solar atmosphere where time-dependent ionization is not important. However, for hydrogen lines formed higher up in the solar atmosphere, HNEI will have severe effects on the line extinctions and source functions. The main effect of HNEI is that the proton density sets the level populations and not the temperature structure. Therefore, structures not reflecting reversed granulation should indicate high ionization regions (leading to increased extinction), showing higher atmospheric layers than reversed granulation. The most prominent features seen sticking out of the reversed granulation background in Fig. \ref{fig:observations_Hepsilon} and Fig. \ref{fig:observations_active_sun} are bright points and dark fibrilar structures. 

\subsection{Bright points}

The bright points in our synthetic and observational \Hepsilon{} images are connected to small magnetic elements. The wide band insert in Fig. \ref{fig:observations_Hepsilon} illustrates that the magnetic structures are connected to \Hepsilon{} bright points.  We explained the formation of \Hepsilon{} above magnetic elements in Sect. \ref{sec:Hepsilon absorption} as the result of density depletion and a different optical depth scale compared to their surroundings. Therefore, we get large \ion{Ca}{ii}~H wing intensities (formed under LTE) with strong \Hepsilon{} absorption lines on top. The strong \Hepsilon{} absorption forms higher in the atmosphere, with a strong decrease of the total source function versus height that is dominated by the \Hepsilon{} source function. If we compare our synthetic \Hepsilon{} profile in Fig. \ref{fig:observational_profiles} with observations of a magnetic element (red profiles), they look remarkably similar. Therefore, we conclude that bright points in \Hepsilon{} observations represent magnetic elements where the \ion{Ca}{ii}~H wing intensities reflect the temperature at the \ion{Ca}{ii}~H wing formation height, and the strong \Hepsilon{} absorption lines reflect an enhanced \Hepsilon{} extinction above magnetic elements where the \Hepsilon{} source function drops rapidly with increasing height.

\subsection{Dark fibrilar structures}

 We could only reproduce a dark fibrilar structure with the Bifrost simulation if we include 3D radiative transfer effects; still, these structures are rare in the synthetic spectra. One example is shown in Fig. \ref{fig:absorption_granule_magnetic}, where the bright fibrilar structure in 1D turns into a (very faint) dark fibrilar structure if 3D effects are taken into account. This feature goes diagonally through the red cross in Fig.~\ref{fig:absorption_granule_magnetic}, extending nearly to the top and right borders of the image. The 3D radiative transfer effects lead to stronger relative absorption contribution above emission (see Fig. \ref{fig:4PFD_x176_x381}) compared to 1D. The reason is the temperature structure of the atmosphere. The temperature structure (or proton densities under HNEI) sets the $n=2$ populations due to the Lyman-$\alpha$ mean radiation field. If we have high-enough temperatures in the lower atmosphere, up to heights where the Lyman-$\alpha$ decouples from the Planck function, it will increase the \Hepsilon{} line extinction. This increases the contribution from higher layers, where the \Hepsilon{} source function is lower than the \ion{Ca}{ii}~H background intensity, thereby giving rise to strong contributions to absorption. Such dark fibrilar structures could in principle be visible in solar observations and not only in synthetic \Hepsilon{} images from simulation. They would mark regions with strong heating: high temperatures that extend from the lower atmosphere up to the lower chromosphere and are not confined to the lower atmosphere.

 It is important to avoid  confusing the Bifrost dark fibrilar structure shown in Fig. \ref{fig:absorption_granule_magnetic} with the dark fibrilar structures observed in more active regions (Fig.~\ref{fig:observations_active_sun}). These fibrilar structures become partly opaque due to increased mass density similar to the \Halpha{} fibrilar chromosphere \citep{Leenaarts2012}. These fibrilar structures outline the chromospheric canopy. There is a clear correlation between the dark fibrilar structures seen in the \ion{Ca}{ii}~H and \Hepsilon{} line core images. On the other hand, the dark fibrilar structure from Bifrost is a result of heating. Dark fibrilar structures similar to those seen in the simulation should be observable in the quiet Sun, but they may be easily confused with other dark structures connected to the reversed granulation background. Dark fibrilar structures should stand out clearly from the \ion{Ca}{ii}~H reversed granulation and not be aligned with the reversed granulation pattern. One example of this is the dark fibrilar structure seen in Fig. \ref{fig:absorption_granule_magnetic} connecting two different granules. It is conceivable that a strong episode of flux emergence may be an event that shows clear signatures of dark fibrilar structures. 

\subsection{\Hepsilon{} in emission}
Lastly, we address locations of \Hepsilon{} emission that mark regions of confined heating in the lower atmosphere (otherwise 3D effects would become important and we would observe a \Hepsilon{} absorption line). Locations with \Hepsilon{} in emission are difficult to identify only by looking at \Hepsilon{} line core images. We have to look at the individual line profiles to identify locations with emission. In Fig. \ref{fig:EN_SE_HNEI}, we outlined \Hepsilon{} emission locations with contours. All locations covered by contours in the simulation are related to temperature enhancements in the lower atmosphere, at heights between $0.3$~Mm and $0.75$~Mm in the Bifrost simulation (see an example in Fig. \ref{fig:4PFD_x176_x381}). We found a significant amount of \Hepsilon{} emission in the Bifrost simulation, but a question arises about whether such emission can also be found in observations covering quiet Sun regions, as \Hepsilon{} is generally thought to be a weak absorption feature in the quiet Sun. To illustrate that the quiet Sun does show regions with emission, we show two example profiles in Fig. \ref{fig:observational_profiles} and compare them to some of our synthetic profiles.

Observing \Hepsilon{} in emission requires high resolution. We interpret the regions with \Hepsilon{} emission as locations marking confined heating events: weak temperature enhancements in the lower solar atmosphere, as illustrated by the formation of \Hepsilon{} emission from the Bifrost simulation.
We are not aware of any spectral line that shows such a clear signature of weak small-scale heating ($\approx 6000$~K) in the lower solar atmosphere. The \ion{Mg}{ii} triplet lines are sensitive to such weak heating events formed at even lower column masses \citep{Pereira2015a} and can serve as a fitting  complement to \Hepsilon{}.

\section{Conclusions}
\label{sec:Conclusion}

We revisited the formation of \Hepsilon{}, inspired by the work of \citet{Ayres1975}, and we propose that the \Hepsilon{} source function is not set by the Balmer radiation field but instead by interlocking via the ground level. We present high-resolution \Hepsilon{} line core images that appear to be dominated by reversed granulation, mostly coming from the \ion{Ca}{ii}~H background radiation as the contribution of \Hepsilon{} is very small under typical quiet Sun conditions. In several cases, we find locations with a different appearance: bright points or dark fibrilar structures that are related to enhanced temperatures and reflect locations where a majority of photons come from the \Hepsilon{} transition. These structures are optically thick to \Hepsilon{} radiation. Therefore, \Hepsilon{} line core images represent a stacked view of different atmospheric layers where the dominant source of photons varies with hydrogen ionization.

Our main findings are as follows:
\begin{itemize}
    \item In most quiet Sun locations, the \Hepsilon{} extinction is minor and the atmospheric layers above the formation height of \ion{Ca}{ii}~H are transparent to \Hepsilon{} radiation.
    \item In magnetic elements and small-scale heating events, the \Hepsilon{} extinction dominates and the atmospheric layers are optically thick to \Hepsilon{} radiation.
    \item Our multi-level source function description suggests that the \Hepsilon{} source function is dominated by interlocking via the term $P_{71} q_{12,7} / P_{21} q_{17,2}$, with the hydrogen ground level as intermediate interlocking level -- and not by the Balmer continuum radiation.
    \item The \Hepsilon{} source function is dominated by the Lyman series via the term $q_{17,2}$, which represents a linear combination of the transition probabilities out of the ground level.
    \item The \Hepsilon{} source function and extinction are both coupled to temperature in the lower solar atmosphere. The extinction via Lyman-$\alpha$, and the source function via the whole Lyman series.
    \item HNEI couples the hydrogen populations to the proton densities, whereas SE couples the population to temperature in the lower solar atmosphere. HNEI has a strong effect on all hydrogen lines, affecting both source functions and extinctions.
    \item We can synthesize \Hepsilon{} in 1.5D with the assumption of HNEI as long as \Hepsilon{} is not formed too high up in the atmosphere, where the 3D effect can become important. 
    \item HNEI has the strongest effect at locations where \Hepsilon{} is formed higher up in the solar atmosphere, such as magnetic elements, dark fibrilar structures, and \Hepsilon{} emission locations. 
    \item \Hepsilon{} emission marks locations of confined heating events in the lower solar atmosphere.
\end{itemize}

The \Hepsilon{} transition is special compared to \Halpha{} and \Hbeta{} because the extinction and source function are coupled to temperature. For \Halpha{} and \Hbeta{}, only the extinctions are coupled to temperature, while their source functions are dominated by scattering, except by strong heating events in the lower atmosphere such as Ellerman bombs \citep{Ellerman1917, Rutten2013}. In \Halpha{} and \Hbeta,{} the line core extinction is high enough that we observe the chromospheric magnetic canopy. For \Hepsilon,{} this is not the case. This gives the unique opportunity for a Balmer series line to look through the magnetic canopy and observe heating events in the lower solar atmosphere. Another interesting fact is that the source function of higher-order Balmer series lines (higher than \Halpha{} and \Hbeta{}) seem to become dominated by interlocking via the Lyman series, and therefore temperature-sensitive. This extinction ``gap'' and temperature sensitivity of the \Hepsilon{} source function opens a new window onto the possibility of observing small-scale weak heating events in the lower atmosphere marked by \Hepsilon{} emission regions, for instance, QSEB reconnection events \citep{Rouppe2016, Joshi2020}, shocks, and others in high spatial and temporal cadence. 
Another positive aspect is that it is relatively easy to observe \Hepsilon{} together with \ion{Ca}{ii}~H. For example, the CHROMIS instrument at the SST can observe both lines without sacrificing too much temporal cadence making the combination of spectral lines a strong diagnostic tool. The \Hepsilon{} and \ion{Ca}{ii}~H combination is also well covered by the ViSP instrument \citep{deWijn2022} at DKIST \citep{Rimmele2020}.

The \Hepsilon{} line is not only of solar interest but also for the wider astrophysics community. Generally, \Hepsilon{} is used to study stellar flare activity \citep{Pavlenko2019}, chromospheric activity in M and K dwarfs \citep{Houdebine2009, Maldonado2017}, active galactic nuclei (AGNs) \citep{Ilic2012}, young active stars \citep{Biazzo2009}, Cepheids \citep{Kovtyukh2015}, and chromospheric active binaries \citep{Montes1995, Montes1996}. \citet{Montes1996} showed a clear surface flux correlation between \ion{Ca}{ii}~H and \Hepsilon{} excess emission for chromospherically active binaries. We can explain this correlation qualitatively. The \ion{Ca}{ii}~H source function for these types of stars is collisionally dominated, formed under LTE conditions, and therefore the intensity is proportional to temperature. Meanwhile, the \Hepsilon{} source function is set by interlocking with a partial sensitivity to temperature and therefore the intensity is coupled to the temperature structure of the atmosphere. The sensitivity of both lines to temperature results in a strong correlation between the two line excess fluxes and explains the steep temperature increase needed to produce \Hepsilon{} emission against the \ion{Ca}{ii}~H background intensity.

We demonstrate that the \Hepsilon{} can be used to detect weak small-scale heating in the lower solar atmosphere at an unprecedented resolution. This could be of particular interest in the studies of QSEBs and it is worthwhile exploring how \Hepsilon{} could help to shed light on these new phenomena in greater detail. Further questions that ought to be considered in future studies include: what higher order Balmer series lines can tell us about QSEBs and whether  stellar \Hepsilon{} emission forms the same way as in the
solar atmosphere as we have predicted, as well as when (and why) the Balmer series lines actually become dominated by interlocking. 
\begin{acknowledgements}

This work has been supported by the Research Council of Norway through its Centers of Excellence scheme, project number 262622, 
and through project number 325491. 
Computational resources have been provided by Sigma2 – the National Infrastructure for High-Performance Computing and Data Storage in Norway. We acknowledge funding support by the European Research Council under ERC Synergy grant agreement No. 810218 (Whole Sun). K. K. acknowledges the insight Mats Carlsson gave during his Stellar Atmospheres II course. 
The Swedish 1-m Solar Telescope is operated on the island of La Palma
by the Institute for Solar Physics of Stockholm University in the
Spanish Observatorio del Roque de los Muchachos of the Instituto de
Astrof{\'\i}sica de Canarias. The Institute for Solar Physics is supported by a grant for research infrastructures of national importance from the Swedish Research Council (registration number 2021-00169).
The Scientific color maps \citep{Crameri_2021} are used in this study to prevent visual distortion of the data and exclusion of readers with colour­vision deficiencies \citep{Crameri_2020}.

\end{acknowledgements}

\bibliographystyle{aa} 

\bibliography{mybib}

\end{document}